\documentclass[iop,apjl,numberedappendix,appendixfloats,floatfix]{emulateapj}
\usepackage{lineno}
\usepackage{color}
\usepackage{amsmath}
\usepackage{url}
\usepackage[figuresright]{rotating}
\usepackage{yfonts}
\usepackage[T1]{fontenc}

\newcommand{\solar}{$_{\odot}$}
\newcommand{\ico}{$I_{\rm CO}$}
\newcommand{\icob}{$I_{{\rm CO},b}$}
\newcommand{\tco}{$^{12}$CO}
\newcommand{\ttco}{$^{13}$CO}
\newcommand{\ceto}{C$^{18}$O}
\newcommand{\htwo}{H$_{2}$}
\newcommand{\joz}{$J$=1$\rightarrow$0}
\newcommand{\jto}{$J$=2$\rightarrow$1}
\newcommand{\kms}{\,km\,s$^{-1}$}
\newcommand{\degree}{$^{\circ}$}
\newcommand{\fdeg}{$^{\circ}$\hspace{-1mm}.}

\newcommand{\vlsr}{$V_{\rm LSR}$}

\def\lapp{\ifmmode\stackrel{<}{_{\sim}}\else$\stackrel{<}{_{\sim}}$\fi}
\def\gapp{\ifmmode\stackrel{>}{_{\sim}}\else$\stackrel{>}{_{\sim}}$\fi}

\newcommand{\lb}{($l$,$b$)}
\newcommand{\lv}{($l$,$V$)}

\newcommand{\uv}{$u_0$,$v_0$}
\newcommand{\uvw}{$u_0$,$v_0$,$w_0$}
\newcommand{\xy}{($x$,$y$)}

\newcommand{\ld}{($l$,$d$)}
\newcommand{\lbd}{($l$,$b$,$d$)}
\newcommand{\sct}{Scutum}
\newcommand{\car}{Carina}

\shorttitle{The CODEX of the Milky Way I. A Ragged Spiral with Midplane Ripples}
\shortauthors{P.\ J.\ Barnes} 

\begin{document}

\title{The Combined Dense Gas and Excitation Atlas of the Milky Way \\ 
I. Mass Distribution and Ripples in the Molecular Layer}


\email{pbarnes@spacescience.org}

\author{Peter J. Barnes}
\affiliation{Space Science Institute, 4765 Walnut St., Suite B, Boulder, CO 80301, USA}




\begin{abstract}
The first components of a new meta-survey, the {\em Combined Dense Gas and Excitation Atlas} (CODEX) of the Milky Way, are presented here with combined data from five previously published surveys: NANTEN+CHaMP, ThrUMMS, FUGIN, and MWISP.  The CODEX is first used to harmonise the combination of parameters used to 
describe the Galaxy's scale and rotation, including $R_0$ and $\Theta_0$ from recent work, since these are critical components of standard techniques for obtaining kinematic distances.  An additional feature is added to these formulae to accommodate local clouds that are otherwise kinematically inconsistent with the Local Standard of Rest.  The structure 
in the combined {\em lV} diagram spanning 230\degree\ of longitude is then analysed, yielding a new Ragged Spiral Arm model that better traces the deprojected \xy\ \tco\ structure.  
The widespread existence of ripples in the molecular layer, first reported in the ThrUMMS data of the Fourth Quadrant (4Q), is confirmed in 
FUGIN data of the First Quadrant (1Q), averaging amplitudes $\sim$40 and $\sim$23\,pc respectively, and wavelength $\sim$4\,kpc in both.  The \xy\ mass distributions as traced by the molecular line data and the published dust extinction data match well across the 4Q+inner1Q, but poorly across the outer 1Q, probably due to 
lower sensitivity in the 1Q extinction data.   In the 1Q+4Q, however, the two $\bar{z}$ distributions do not match: 
the dust ripple amplitude is $\sim$3$\times$ smaller than, and opposite in sign to, the gas ripples.  This is nevertheless consistent since the dust extinction is insensitive to the densest molecular gas. 
\end{abstract}



\section{Introduction}

Defining the structure of the Milky Way has been an enduring challenge for well over two centuries, since \citet{h17} first described its flattened aspect.  Its true size, however, was not really apprehended until more than a century later, when \citet{s18} first mapped the extent of the globular cluster system around the Galaxy.  Clues to the spiral nature of the disk, and especially the configuration of the cold interstellar medium (ISM), had to await the advent of radio astronomy and maps of the HI distribution \citep[e.g.,][]{v65} and molecular gas \citep[primarily traced by CO;][]{hd15}.  Thus, much observational work on the Milky Way's large-scale ISM structure has been pursued at radio wavelengths.  This progress has also generated theoretical challenges, most intractably (i.e., for the last $\sim$70 years) on the origin of spiral structure itself, as extensively reviewed by \citet{sm22}. 

Equivalent work on stellar Population I components has historically been more challenging, due to the heavy optical obscuration in the Galactic Plane, limiting our ability to map stars and young clusters in the disk beyond a few kiloparsecs (kpc).  Despite the advantages offered by infrared data (such as from {\em Spitzer}) that better penetrate the dust, some basic puzzles remain unsolved even today.  For example, whether the Milky Way is a two- or four-armed spiral has been debated for 70+ years \citep[see the excellent summary by][their Appendix A]{k21}, continuing with the most recent {\em Gaia} results, which seem to favour a muted, multi-arm pattern in the young stellar disc \citep{g23}.

Since $\sim$2005 mm-wave molecular line mapping has enjoyed a huge increase in productivity, due to the development of advanced, multiplexed receiver systems that allow simultaneous mapping of multiple species in molecular clouds.  There are now on the order of 20 molecular line surveys across various sectors of the Galactic Plane, each typically covering 20--120\,deg$^{2}$ in area. 
While all are reasonably sensitive to the mapped line emission, with similar sub-km/s velocity resolution and arcminute-scale angular resolution that translates to a useful pc-scale physical resolution for the typical giant molecular cloud (GMC) at 3\,kpc distance, these surveys are not equal.  Each survey team has also engaged in a wide range of scientific exploits, which are individually interesting but not always directly comparable, one example being the disparate methods used to create ``cloud catalogues.'' 

To address some of the larger issues of Galactic structure and the global properties of molecular clouds, it would be desirable if several of these newer molecular line surveys could be harmonised and combined onto a single grid or database.  A first attempt at doing this (with 5 of these surveys, as described below, plus crosschecks with a sixth) is introduced here as the {\em Combined Dense Gas and Excitation Atlas} of the Milky Way, or ``CODEX.''\footnote{A codex is defined as an ancient manuscript text in book form, or an official list of medicines, chemicals, etc \citep[Dictionary v2.3.0,][]{apple23}.\vspace{1mm}}  Such collected data would simplify analysis of several whole-Galaxy issues, like its spiral structure as measured in various components, widely-used rotation models that enable kinematic distance solutions, or how different subpopulations of stars, clouds, or implied fields (like the Galactic potential) are all related to each other in three dimensions.  Being a compilation of dense gas tracers, the CODEX is complementary to the {\em Gaia} mission, which has already enabled numerous holistic studies of large fractions of the Galaxy's stellar component at optical wavelengths \citep[e.g.,][]{g21,g23}.  The soon-to-be launched ($\sim$late 2026) {\em Roman Space Telescope} promises to do likewise in the near-IR \citep{rgps}, thanks to its widefield imaging capability that will be able to penetrate more deeply through a large portion of the inner Galactic Plane than {\em Gaia}, and at much higher astrometric accuracy than {\em Spitzer}.

This paper is organised as follows.  \S\ref{assemble} describes how the various surveys' data are brought together so as to permit a holistic analysis of Galaxy-wide parameters, including those describing the Milky Way's overall structure and those of molecular cloud properties as a function of position across the Milky Way.  \S\ref{rotparsumm} \& \S\ref{rotmodels} summarise recent work on defining the Milky Way's size and rotation parameters, and how the CODEX can improve upon these results by comparing and combining information from multiple surveys. 
In \S\ref{spiral} these updated parameters are applied to produce new \xy\ deprojections of the \tco\ survey data, 
including the integrated intensity \icob, mean latitude $b$, and mean height $z$ distributions.  A new ``Ragged Spiral Arm'' model is fitted to the mass distribution, and the midplane deviations presented.  These 3D gas distributions are then compared with the 3D dust distribution in \S\ref{dust}.  The conclusions appear in \S\ref{concl}.

\section{Assembling the CODEX}\label{assemble}

The CODEX is intended to provide a platform for generating a set of regularised data products from disparate original sources, in order to facilitate further analysis.  One such useful application is to provide composite or consistent moment maps (or other derived quantities) from the merged public data cubes of various surveys.  These cubes will typically be in the \joz\ or \jto\ lines of \tco, \ttco, and/or \ceto, and could (for example) comprise multiple surveys in a particular line with wider longitude coverage than in a single survey, or analysis of multiple lines from different surveys in a single, focused \lb\ range, depending upon the scientific objective.

The initial version of the CODEX described here combines the public \tco\ \joz\ data from five surveys: cubes from FUGIN \citep{um17}\footnote{http://jvo.nao.ac.jp/portal/nobeyama/fugin.do}, ThrUMMS \citep{b25}\footnote{https://doi.org/10.26131/IRSA628}, and NANTEN \citep{y05} (hereafter collectively referred to as FTN), moment maps from MWISP \citep[cubes were not available at the time of writing;][]{s20}\footnote{https://cade.irap.omp.eu/dokuwiki/doku.php?id=mwisp} (collectively, WFTN), plus distance information on molecular cloud complexes from CHaMP \citep[which are contained within the NANTEN sky coverage;][]{b11}\footnote{https://doi.org/10.26131/IRSA542}.  The results are then compared with similar CfA survey data \citep{dht01}\footnote{https://lweb.cfa.harvard.edu/rtdc/CO/}.  All data were obtained from their respective websites and/or data repositories as standard FITS files, and then imported into \textsc{Miriad} \citep{s95} for processing using custom Unix shellscripts.  For this paper, a standard set of moment maps was computed for each cube using the SAM algorithm \citep{r90,b15}, which provides better noise filtering and mitigation than standard moment analysis.\footnote{Conventionally, 0th moment maps are simply maps of emission integrals, whether over velocity \vlsr\ (yielding maps of integrated emission across the sky), latitude $b$, or height $z$ (yielding latitude- or height-integrated emission over \lv\ or \xy\ space).  Similarly, 1st moments are intensity-weighted mean velocities/latitudes/heights over their respective spaces; and 2nd moments are maps of dispersions around the respective means.} 

Not coincidentally, the FTN surveys cover almost exactly the same longitude range of the planned {\em Roman Galactic Plane Survey} or RGPS \citep[$l$ = 286\degree--0\degree--51\degree;][]{rgps}.  Both run between the two major spiral arm tangencies in Carina and Sagittarius (traditionally thought of as tracing the Sgr-Car Arm) of the inner Galaxy, covering ~80\% of the Milky Way's mass.  This combination, in addition to other surveys in the same longitude range like CHaMP \citep{b11} and SEDIGISM \citep{sed21}, should prove very fruitful for future {\em Roman} science.

\section{Revised Scale and Rotation Parameters}\label{rotparsumm}

The primary diagnostic used to infer the Galaxy's overall scale and rotation parameters is the composite \lv\ zeroth-moment diagram (i.e., integrated intensity across all $b$).  This is obtained by mosaicking the individual SAMed \lv\ diagrams from the original FTN \tco\ data cubes, plus the original MWISP \lv\ diagram (but here noise-masked to approximate the SAM filtering), onto a common \lv\ grid (from ThrUMMS), with 24$''$ pixels in $l$ and $\sim$0.09\,\kms\ channels.  This slightly underresolves the spatial information in the other surveys, but this is not an issue for the wide field analysis discussed herein, and ThrUMMS' higher velocity resolution is preferred to avoid loss of information for the kinematic analysis.  
The \lv\ diagrams then form the principal constraint for models of Galactic rotation and their various parameters.

A great deal of work has been published on this modelling since the early days of HI observations in the 1950s, but reviewing that literature is well beyond the scope of this paper.  Here we limit ourselves to describing some recent work in this area, and offer a few updates.  These details are given in {\color{red}Appendix \ref{rotmodels}}: in summary, we use the BeSSeL solutions (\uvw) for Solar Motion \citep{r19} but with the latest values for the Galactic scale parameters $R_{0}$ and $\Theta_{0}$ from VERA \citep{o24}.  We propose an additional feature in the standard kinematic distance formulae that can take into account deviations from pure rotational motion over small portions of the Galactic disk.  As an example of such a Regional Standard of Rest (RSR), we apply this formalism to non-rotational motions of some low-mass clouds near the Sun, which are otherwise at ``forbidden'' velocities compared to the Local Standard of Rest (LSR), and thus do not have solvable kinematic distances.

\begin{figure}
	\includegraphics[angle=180,width=\columnwidth]{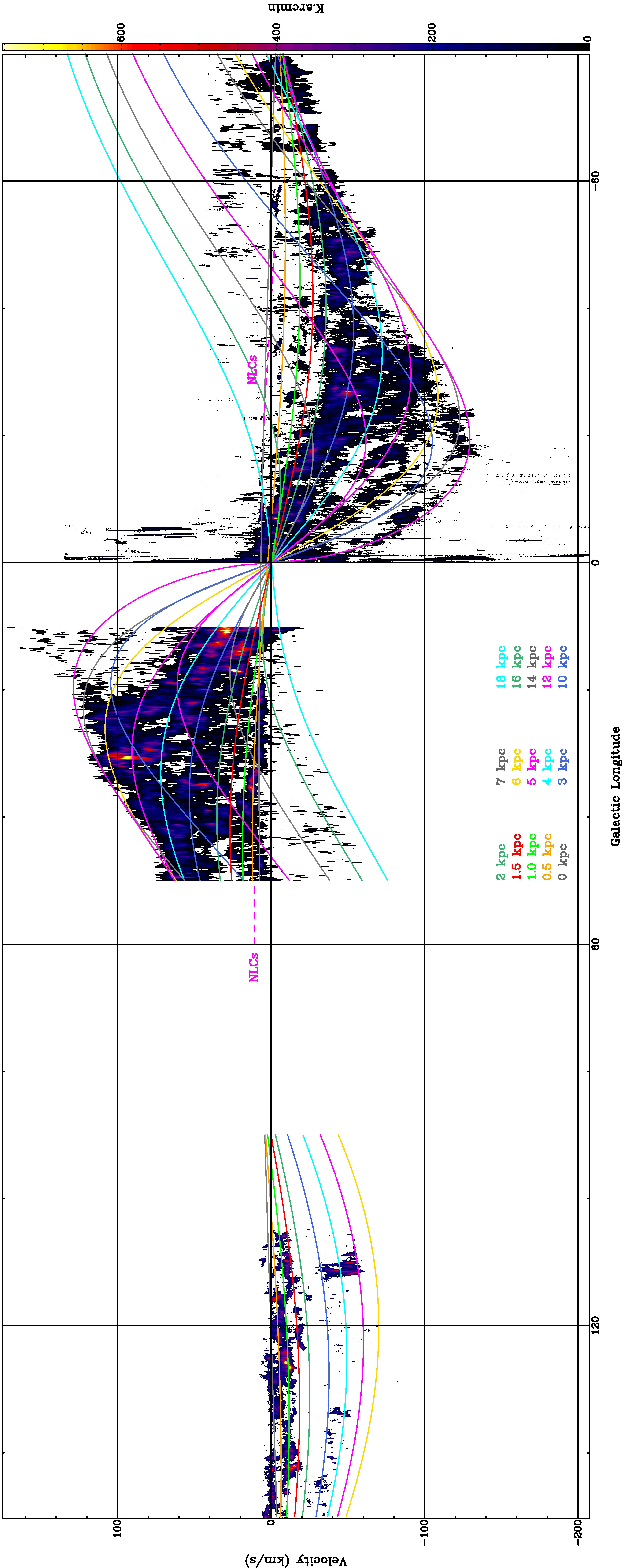} 

	{\color{blue}\vspace{-190mm}\hspace{54mm}MWISP \\

			\vspace{68mm}\hspace{73mm}FUGIN \\

			\vspace{55mm}\hspace{56mm}ThrUMMS \\

			\vspace{29mm}\hspace{25mm}NANTEN \\
	} 

	\vspace{13mm}
	\caption{\footnotesize Same \lv\ diagram for all WFTN data as in Fig.\,\ref{oldLVmodels}, but on a more muted colour scale and overlaid by 
		contours of constant heliocentric distance (as labelled) for the BVTFg rotation model described in the text, plus its tangent velocity curve 
		(magenta).  Note the difference between the \vlsr\ = 0 curve shown here and that in Fig.\,\ref{oldLVmodels}: close to this lie the narrow-line 
		clouds described in the text, indicated by a dashed magenta curve labelled NLCs.}
	\label{newhelio}
\end{figure}

\begin{figure}
	\centering{\includegraphics[angle=180,width=\columnwidth]{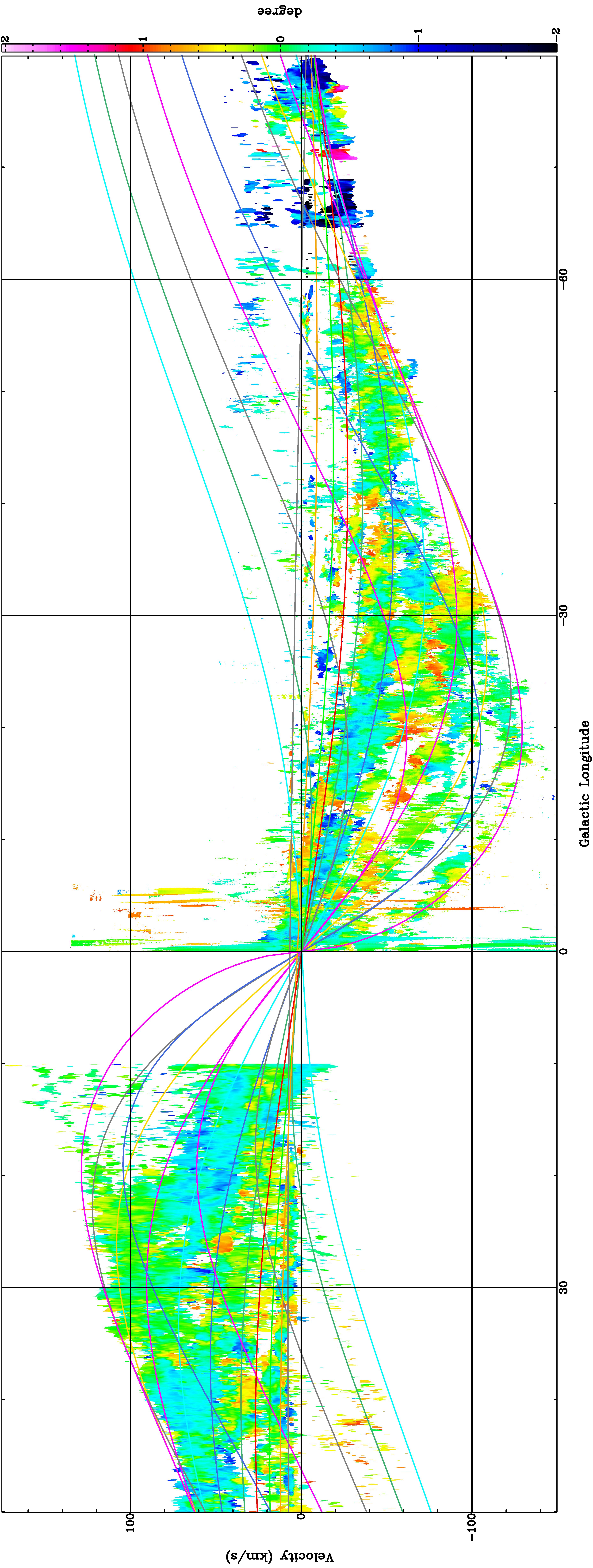}}

	\vspace{-1mm}
	\caption{\footnotesize First latitude-moment (intensity-weighted mean latitude) in \tco\ data from the FTN surveys, a subset of the longitude 
		coverage in Fig.\,\ref{newhelio}.  Overlaid are the same heliocentric contours for the BVTFg rotation model.}
	\label{ftn1lv}
\end{figure}

\begin{figure}
	\centering{\includegraphics[angle=180,width=\columnwidth]{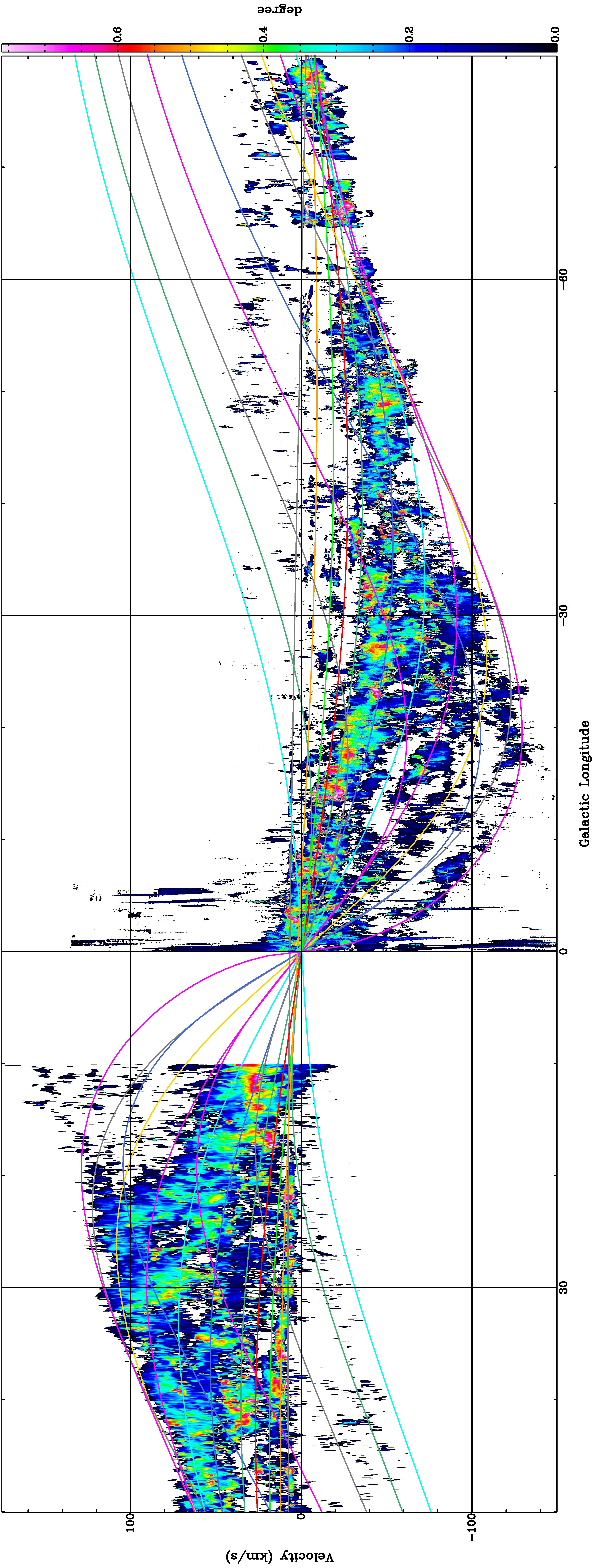}}

	\vspace{-1mm}
	\caption{\footnotesize Second latitude-moment (intensity-weighted latitude dispersion or thickness) in \tco\ data from the FTN surveys, 
		similar to Fig.\,\ref{ftn1lv}, including the BVTFg heliocentric contours and tangent curve.}
	\label{ftn2lv}
\end{figure}

We use this new, most-preferred ``BVTFg'' model (which is {\bf identical} to a BeSSeL+VERA-only model for distances $d$ $>$ 1\,kpc) to produce kinematic deprojections of the \lv\ data from the surveys described in \S\ref{assemble}.  The new heliocentric solutions can be plotted for various $d$ on \lv\ diagrams for this model, as in {\color{red}Figures \ref{newhelio}--\ref{ftn2lv}}.  We supply \textsc{SuperMongo} \citep{lm00} code to quickly generate these heliocentric contours (and related plots) via the GitHub URL in {\color{red}Appendix \ref{code}}.  To obtain a distance from \lv\ data that takes into account the scaling factor $s$ and the standard LSR correction, one simply inverts the calculation with the usual kinematic formulae.  Practically, however, transforming an entire 2D or 3D dataset to a distance- or \xy-based coordinate system (e.g., as in \S\ref{spiral}) is computationally intensive.

\section{Spiral Structure as Traced by \tco}\label{spiral}
\subsection{A Distorted Spiral Arm in the XY Plane}\label{zoom}

\begin{figure*}
	\includegraphics[angle=-90,width=179mm]{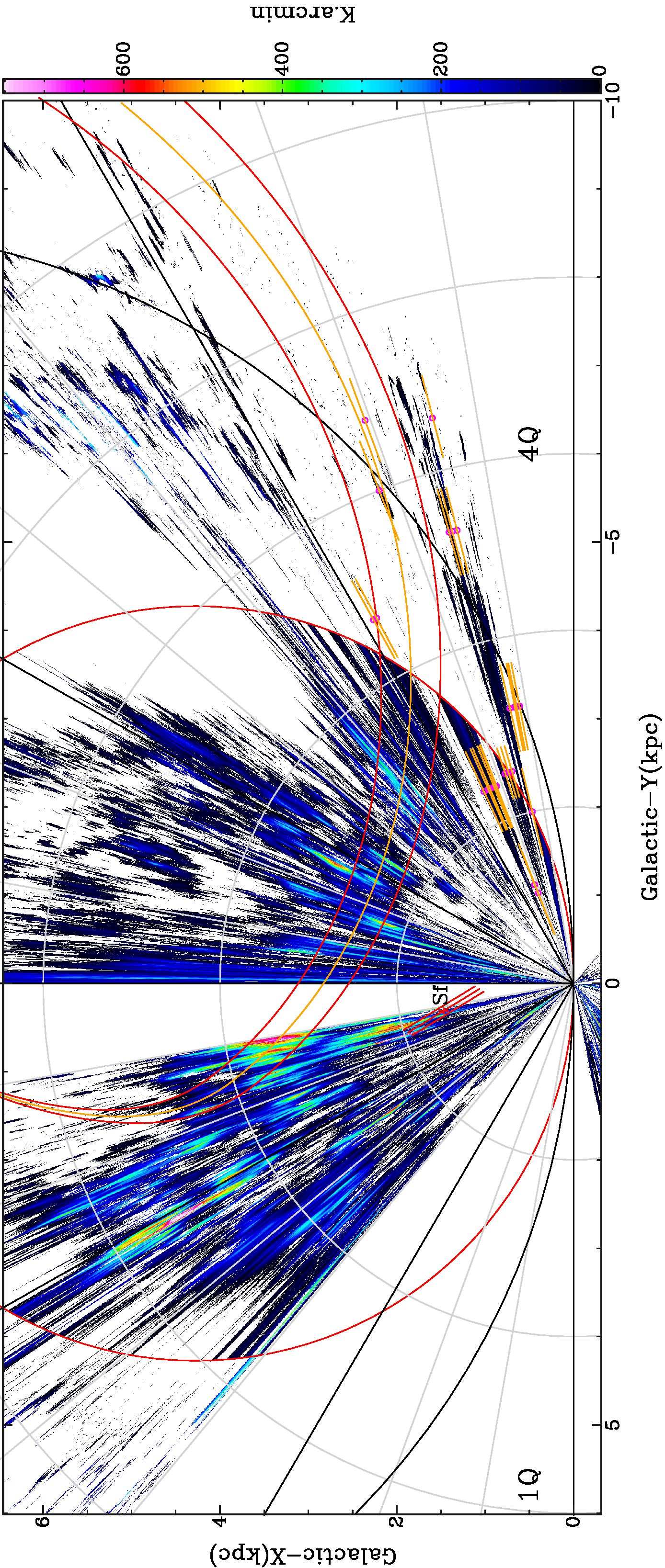}		

	\vspace{-5.2mm}
	\includegraphics[angle=-90,width=179mm]{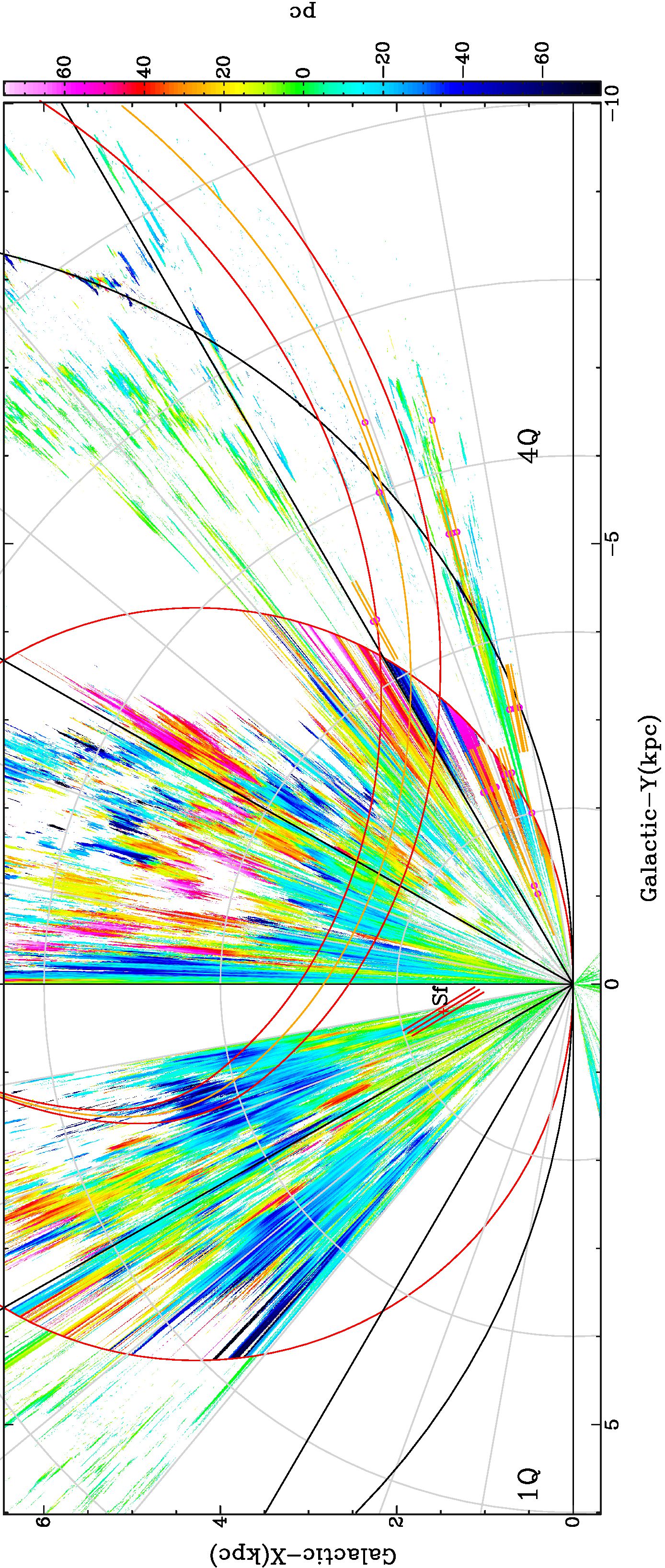}

	\vspace{-5.2mm}
	\includegraphics[angle=-90,width=179mm]{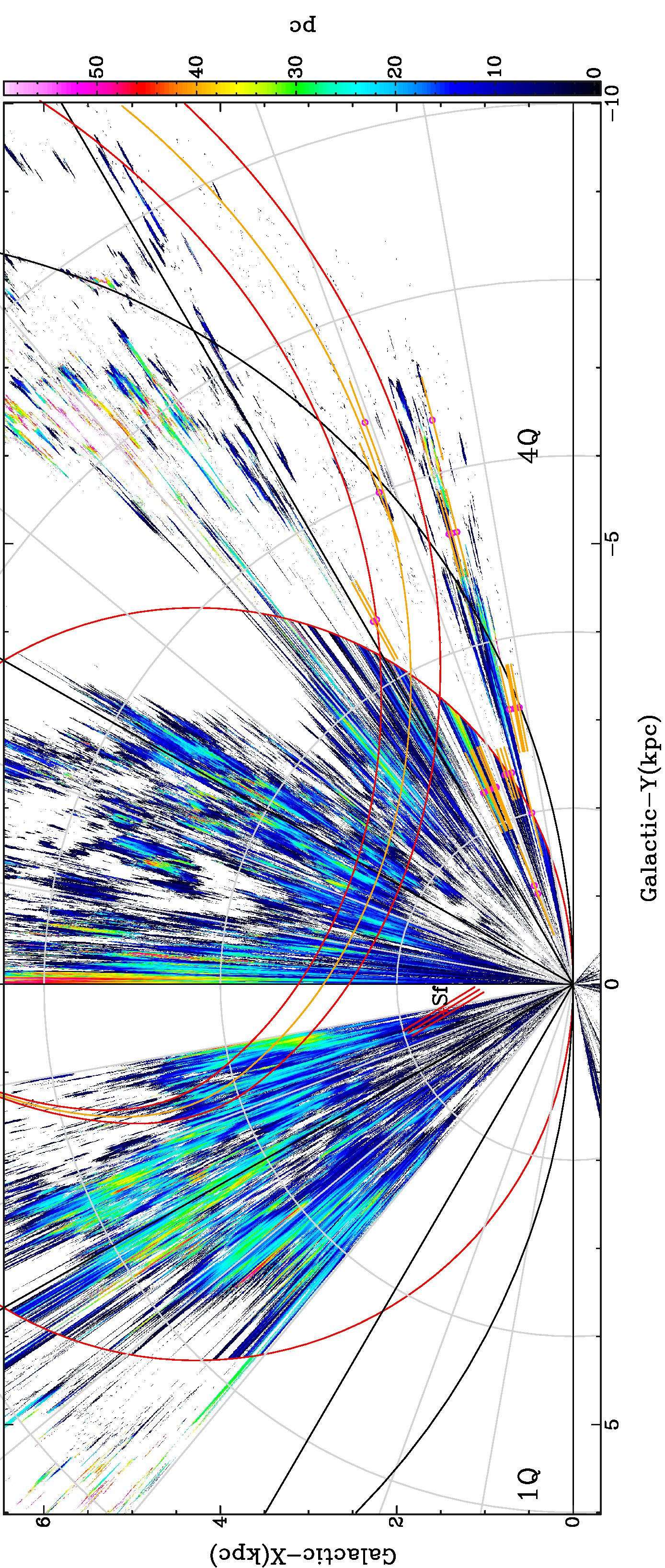} \\
	{\color{blue}\footnotesize

	\vspace{-160mm}\hspace{132mm}Integrated \tco\ emission \\

	\vspace{65mm}\hspace{132mm}Mean height of gas layer \\

	\vspace{65mm}\hspace{132mm}Thickness of gas layer \\
	} 

	\vspace{-219mm}
	\hspace{-1mm}\includegraphics[angle=0,scale=0.266]{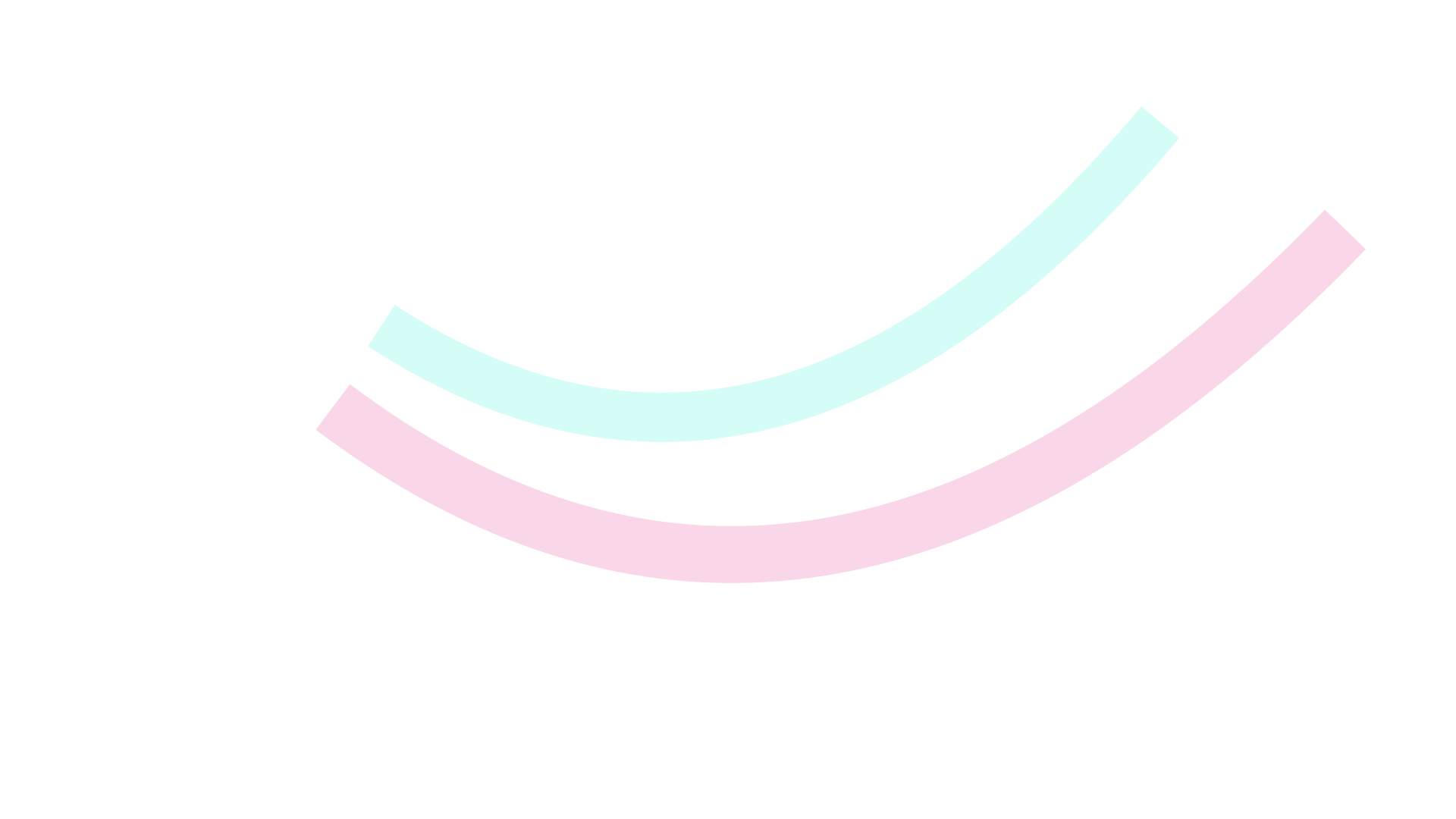} \\

	\vspace{-33.6mm}
	\hspace{-1mm}\includegraphics[angle=0,scale=0.266]{shaded-Reid-arms.png} \\

	\vspace{-33.6mm}
	\hspace{-1mm}\includegraphics[angle=0,scale=0.266]{shaded-Reid-arms.png} \\

	\vspace{-17mm}
	\caption{\footnotesize Same \tco-moment data as in Figs.\,\ref{oldLVmodels}--\ref{ftn2lv}, i.e., latitude-moments 0, 1, and 2 from top to bottom, but 
		with observed \lv\ coordinates converted to an \xy\ grid --- where the Sun is at (0,0) and the Galactic Centre at ($R_0$,0) --- via the BVTFg 
		rotation model described in the text, and the $\zeta^{+}$ near/far distance filter of \citet{b25} applied.  These panels are zoomed in from the 
		full transformations (Figs.\,\ref{wftn-xy0}--\ref{ftn-xy2}) to show more detail in the inner Galaxy.  Overlaid features are: a heliocentric polar 
		coordinate grid (every 2\,kpc in $d$ and 10\degree\ in $l$ shown in grey, every 30\degree\ in black), with quadrant labels also in black; 
		the galactocentric solar circle at $R_0$ (black); 
		the kinematic tangent circle (red); spiral arm models in blue (Sct-Cen) and magenta (Sgr-Car) shading \citep{r19} and red+orange curves 
		(RSA); the Sgr feather near \xy\ = (1.5,0.5)\,kpc (short red lines labelled ``Sf'') described by \citet{k21}; and 20 CHaMP Regions within 
		300\degree\ $>$ $l$ $>$ 280\degree\ (small magenta circles with orange distance errorbars) from \citet{b11}.
	}\label{xysmall}
\end{figure*}

Having decided upon reasonable Galactic scale and Solar Motion parameters, the next step is to derive \ld\ or \xy\ solutions for all the \lv\ data in Figures \ref{oldLVmodels} and \ref{newhelio}, with a standard kinematic distance calculation that includes inverting Eq.\,(A1).  Here the \xy\ Cartesian coordinate system is a heliocentric one, where the Sun is at (0,0) and the Galactic Centre is at ($R_0$,0).  The full-scale Cartesian versions of these for all available WFTN data are shown in {\color{red}Appendix \ref{wideview}}; \ld\ versions were also computed but, since they contain essentially the same information as the \xy\ deprojections, are not shown here.  These full-scale images are quite large, $>$5000$^{2}$ pixels, 
so {\color{red}Figure \ref{xysmall}} presents zoomed-in versions of these diagrams which are more useful to a discussion of structure in the inner Galaxy.

The main question is how well (or not) the pattern of peak latitude-integrated \tco\ intensity,\footnote{The \tco\ surveys' native units are brightness temperature, $T_{b}$, but for notational convenience we use intensity $I$, since they are directly related by the Planck Law $I$ = 2$kT_{b}/\lambda^{2}$.} i.e. \icob\ = $\int I$($^{12}$CO)\,d$b$ (Fig.\,\ref{xysmall}, top panel), follows any spiral arm model, whether from the literature or newly-devised.  To this end, all panels of Figure \ref{xysmall} 
also show overlays of the \cite{r19} Scutum-Centaurus and Sagittarius-Carina Arms (hereafter Scutum and Carina Arms for brevity), preserving their colour scheme as blue- or magenta-shaded curves respectively.  Their widths represent each Arm's extents of $\pm$4.7\% in radius $r$.  For this study, however, both Arms have been scaled up in galactocentric radius by the ratio of the adopted $R_{0}$ in this work to that of \cite{r19}, i.e, 8.55/8.15 = 1.049.  Without this adjustment, the \sct\ and \car\ Arms follow the \tco\ distribution even more poorly than described below, as noted by \cite{z25} (see also \S\ref{dust}).

The \cite{r19} spiral arm models are the product of a major VLBI effort to obtain precise maser astrometry in massive star-forming regions, i.e., molecular clouds, in order to better define the structure of the Milky Way.  However, they have had limited access to the southern Galaxy due to the smaller number of cm-wave antennas in the Earth's southern hemisphere.  Therefore it should not be too surprising that the spiral arm models, based mostly on northern-hemisphere data, do not reproduce well the Milky Way's observed spiral structure in the south.  Supported by the dust extinction results, the kinematic modelling of \tco\ data now gives us a tool to derive an improved model of the Milky Way's dense gas spiral pattern.  While the \tco\ data may lack the precision of the VLBI data, they make up for that in volume and coverage.

Looking at Figure \ref{xysmall}, in the vicinity of the \sct\ Arm and in the sense of decreasing galactocentric longitude $\beta$, there is a ridgeline of highest \icob\ across the 1Q+4Q data.  This ridgeline first peaks near \xy\ $\approx$ (5.0,2.2)\,kpc, but at a similar $\beta$ there is a higher peak $\approx$ (4.4,2.6).  Within the 1Q data, the ridgeline more generally runs from $\approx$ (5.1,3.0) to (3.3,0.6).  Compared to the \cite{r19} models, this ridgeline starts close to the \sct\ Arm at $l$ $\sim$ 30\degree, and roughly (but not very closely) follows the \sct\ Arm to $l$ = 10\degree.  In the 4Q data, the main ridgeline continues to track with the \sct\ Arm from $l$ = 360\degree--330\degree, but does so only very approximately, with large deviations to $\sim$1\,kpc inside and $\sim$2\,kpc outside the \sct\ Arm.  The ridgeline then meanders out towards the \car\ Arm from $l$ $\approx$ 325\degree\ to 300\degree.  As modelled, the \sct\ Arm curves inward of this meander to a significant degree, about 2\,kpc inside the ridgeline by the time the \sct\ Arm reaches the tangent circle.  
Overall, the high-\icob\ ridgeline is oriented at a distinctly higher pitch angle ($\sim$35\degree) than \cite{r19}'s \sct\ Arm (14\degree) across much of the 1+4Q (40\degree\ $>$ $l$ $>$ 300\degree).

By comparison, the \car\ Arm is not really well-defined in the \tco\ data: rather, there are a number of disjointed concentrations and large voids along the \cite{r19} \car\ Arm curve.  For example, slightly beyond the head of the \car\ Arm at \xy\ $\approx$ (2.75,2.6), a diffuse feature follows the \car\ Arm orientation along $l$ $\approx$ 35\degree\ for 1--2\,kpc, but then peters out within $d$ $\sim$ 3\,kpc.   Then from $l$ = 30\degree\ to 295\degree, there is hardly any \tco\ emission along the nominal \car\ Arm, except for the rather compact Sgr feather.\footnote{The BVTFg model places the Sgr feather at \xy\ $\approx$ (2.0,0.5), about 0.5\,kpc further from the Sun than its known location \citep{k21}.  This actually suggests that the gaussian taper used in Eq.\,(2) should have a scale $d_{s}$ closer to 2.5\,kpc than 0.5\,kpc, which would move the Sgr feather and most other local structures ($d$ \lapp\ 2\,kpc) about 0.5\,kpc closer to the Sun, without changing the larger-scale features described here.  This was {\em not} done in the BVTFg model (i.e., $d_{s}$ was left at 0.5\,kpc), pending further evidence to support it.}  Other concentrations aligning somewhat with the \car\ Arm lie at $d$ $\sim$ 2--3\,kpc near $l$ $\sim$ 295\degree\ and 285,\hspace{-1mm}\degree\ but the pattern of NANTEN data in 300\degree\ $>$ $l$ $>$ 280\degree\ is very patchy.  Also in these directions, the kinematic solutions suffer from gross ``velocity-smearing,'' so it isn't clear how well the NANTEN data actually trace the \car\ Arm.  The mostly classically-determined distances of the CHaMP Regions would seem to confirm something like the \car\ Arm does exist, but only starting from the tangent circle and going perhaps 5\,kpc outwards from there.  One could just as easily argue these clouds are better described by a few discrete, large-scale feathers coming off the main part of the \sct\ Arm.

The overall impression is that the two \cite{r19} Arms do not trace out the ridgeline of the \tco\ distribution that well, particularly for the \car\ Arm, but also the \sct\ Arm to some extent.  Indeed, one can argue that this single high concentration of molecular gas in the 1+4Q comprises one rather ragged manifestation of a single spiral arm to come off the Milky Way's Bar, one with a very high gas content and star-formation activity.  While many studies have found spiral arm segments at various locations throughout the stellar disk, and a pattern that is often suggestive of a multi-armed spiral \citep[e.g.,][]{g23,dc26}, the point here is that {\em the dense molecular gas} that traces the most vigorous star formation seems not to be so arranged in the \tco\ survey data.

However, a counter-argument to this picture can be made.  With \cite{r19}'s Carina Arm approaching the Sun to almost $d$$\sim$1\,kpc, it is possible that the FUGIN+ThrUMMS latitude extent of $\pm$1\degree\ has missed some of the nearer portions of this Arm, which in projection, could extend to higher latitudes ($b$ = 1\degree\ at 1\,kpc corresponds to a height of only $z$ = 17\,pc).  Fortunately, we can test this idea with the \tco\ data of the CfA survey \citep{dht01}, which although of lower resolution ($\sim$8$'$ where fully sampled), has a much broader latitude extent of at least $\pm$5\degree\ (and even broader at some longitudes).

{\color{red}Appendix \ref{cfaviews}} shows moment maps of the CfA data, computed similarly to those in Appendix \ref{wideview}, i.e., deprojected to \xy\ and disambiguated with the $\zeta$ function as described by \cite{b25}.  In particular, the integrated intensity map of {\color{red}Figure \ref{cfa-xy0}} is instructive.  Significant additions to the molecular cloud distribution (compared to {\color{red}Fig.\,\ref{wftn-xy0}}) include (1) a feature at $d$$\approx$1.6\,kpc and $l$$\approx$40\degree, just outside the nominal Carina Arm extent; (2) an even brighter collection of features at $d$$<$0.5\,kpc extending across $l$$\approx$20\degree--40\degree, but $\sim$1\,kpc beyond the Carina Arm outer edge; and (3) two other bright features along $l$$\approx$15\degree--0\degree--345\degree\ and 80\degree\ at larger (but kinematically very uncertain) distances $d$$\sim$2--5\,kpc, which even so are not aligned with the Carina Arm.  Besides these bright features, there is weaker emission filling in some of the gaps in the nominal Carina Arm across the 4Q, e.g., fleshing out some structure around the NANTEN/CHaMP complexes described above.  But overall, these additions do not help to create an impression of a contiguous structure aligned with the \cite{r19} Sgr-Car model.

Instead, the overall impact of Figure \ref{cfa-xy0} is to underscore the idea that, {\bf{\em in the molecular gas}}, the \xy\ structure of the dense gas layer tends to be dominated by single, rather ragged, high pitch angle structure between the Sun and the Galactic Centre, which historically corresponds to the ``Molecular Ring'' of the first CO surveys from the 1980s.  The large pitch angle manifest in all of Figures \ref{xysmall}, \ref{wftn-xy0}, and \ref{cfa-xy0} is a direct consequence of this distribution's asymmetric structure in the corresponding $lV$ diagrams, combined with the latest rotation model parameters \citep[essentially the BeSSeL \uvw\ solutions, as modified by the VERA results:][]{r19,o24}.

\begin{figure}
	\hspace{-1mm}\includegraphics[angle=-90,width=87mm]{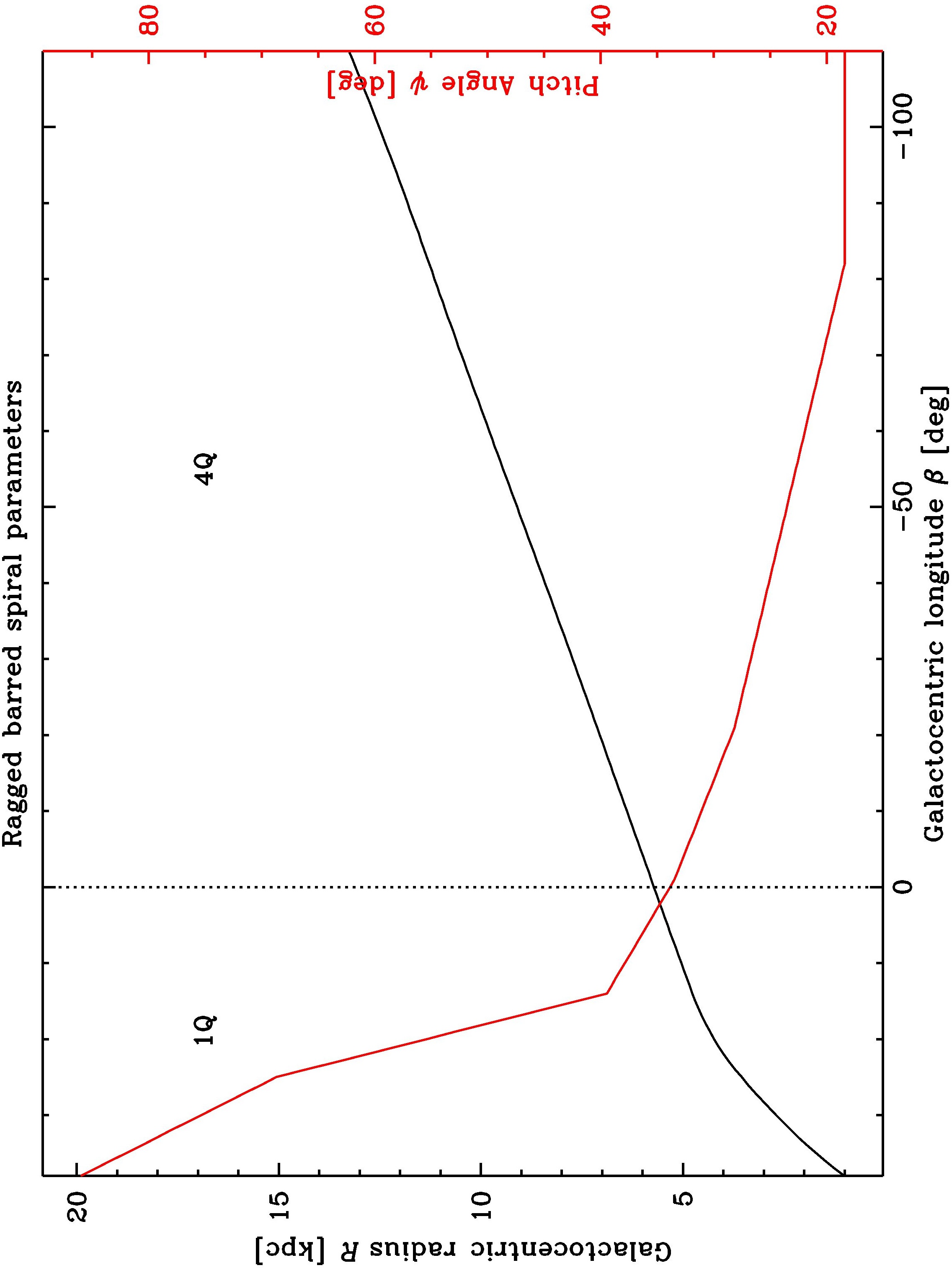}
	\caption{\footnotesize Parameters of the Ragged Spiral Arm shown in Figs.\,\ref{xysmall}, \ref{smalldust}, \ref{zdiff}, and \ref{wftn-xy0}--\ref{wftn-dust}. 
		The Arm's galactocentric radius $R$ (black) and pitch angle $\psi$ (red) are both functions of the Arm's galactocentric longitude 
		$\beta$.  The pitch angle $\psi$ was varied as 
		needed to best match the mass distribution in Figs.\,\ref{xysmall} and \ref{wftn-xy0}, but kept a continuous function of $\beta$ 
		in order to preserve, as far as possible, a smooth appearance for $R$($\beta$) in the deprojected maps.  The vertices \citep[or 
		``kinks'' in the sense of][]{r19} and gradients of the red $\psi(\beta)$ function are also listed in Table \ref{rbs}.
	}\label{barmodels}
\end{figure}

That such a spiral arm is better-defined in the 1Q where the pitch angle $\psi$ is decreasing rapidly from $\sim$80\degree to $\sim$20\degree, and more poorly-defined in the 4Q with a number of gas concentrations away from the ridgeline, should not be very surprising.  But it means that the long-standing argument about whether the inner Galaxy manifests a 2-arm or 4-arm structure in the overall stellar distribution \citep{k21} seems to be somewhat beside the point, as far the dense molecular gas is concerned.  Here, the pattern is grossly dominated by a single, rather ragged spiral arm between us and Sgr A*.  Outside the Solar Circle, several molecular gas surveys reveal both a Local Arm (also called the Local Spur or Orion Arm; see {\color{red}Fig.\,\ref{cfa-xy0}}), and the much weaker but very clear signature of the Perseus Arm.  In combination, these features can also approximate multi-arm models, but for this work, we simply point out the dominance of the high pitch-angle arm/Molecular Ring on the near side of the inner Galaxy.

\subsection{A New Spiral Arm Model}\label{model}

Based on the high pitch angle distribution of the molecular gas, one can define a completely new Ragged Spiral Arm (hereafter RSA) model.  This is shown in Figures \ref{xysmall} as an orange plus 2 red curves (indicating a midline and a half-width of 5\% in $r$ as before) with the parameters given in {\color{red}Figure \ref{barmodels}} and {\color{red}Table \ref{rbs}}.  Where they lie close together across the 1Q and $\sim$1/3 of the 4Q, the RSA roughly matches the \sct\ Arm in position, but has a much higher pitch angle.  It then veers outward towards the \car\ Arm in order to track the deviation of the ridgeline towards larger $r$, from $l$ $\sim$ 330\degree.  Thereafter, the RSA approaches the \car\ Arm, but beyond a distance $d$ $\sim$ 5\,kpc, none of the three Arm models are well-constrained by the \tco\ data.

\begin{deluxetable}{cccccc} 
\tabletypesize{\footnotesize}
\tablecaption{Parameters of the Ragged Spiral Arm}
\tablehead{
	\colhead{$\beta$ range} & \colhead{$R_{\rm start}$} & \colhead{$R_{\rm end}$} & \colhead{$\psi_{\rm start}$} & \colhead{$\psi_{\rm end}$} & 
	\colhead{$\Delta\psi$\tablenotemark{$^{a}$}}
	}
\startdata
	38\degree\ to 25\degree & 1.00\,kpc & 3.55\,kpc & 86\degree & 68\fdeg70 & --1\fdeg331  \\
	25\degree\ to 14\degree & 3.55 & 4.77 & 68\fdeg70 & 39\fdeg45 & --2\fdeg659  \\
	14\degree\ to   --1\degree & 4.77 & 5.79 & 39\fdeg45 & 33\fdeg47 & --0\fdeg399  \\
	--1\degree\ to --21\degree & 5.79 & 7.13 & 33\fdeg47 & 28\fdeg14 & --0\fdeg266  \\
	--21\degree\ to --82\degree & 7.13 & 11.28 & 28\fdeg14 & 18\fdeg41 & --0\fdeg160  \\
	--82\degree\ to --180\degree & 11.28 & 19.9 & 18\fdeg41 & 18\fdeg41 & 0\degree \\
\enddata \vspace{-3mm}
\tablenotetext{a}{This is the change in pitch angle $\psi$ per degree change in galactocentric longitude $\beta$ ($\beta$ is the independent variable 
	in this construction).  The arm starts at galactocentric coordinates ($R$,$\beta$) = (1\,kpc, 38\degree) and, for each 1\degree\ decrement in 
	$\beta$, builds out in $R$ and $\psi$ with the given $\Delta\psi$.}
	\label{rbs}
\end{deluxetable}

This new RSA model is not meant to be definitive, but rather, illustrative--- many similar models can be constructed with slightly different positions and orientations.  The point is really to show that a classical barred spiral arm, with a strongly decreasing $\psi$ from the bar to the main part of the disk, is a tenable fit to the data.  For example, the RSA model starts at $\beta$ = 38\degree, and doesn't really reach the position of the start of the ridgeline at \xy\ $\sim$ (5,3).  One could just as easily construct a RSA model that starts at (say) $\beta$ = 50\degree\ and then has a sharper turn at $\sim$(4.5,2) to follow the ridgeline from that point.  The smaller $\beta_{0}$ starting value in Table \ref{rbs} was chosen instead to pay some heed to current bar data and models that suggest it is oriented roughly at $\beta$ $\approx$ 20\degree\ \citep{g23}, which the illustrated RSA reaches at \xy\ = (4.6,1.4).  Also, many barred spirals show that vigorous star formation activity can occur somewhat ``downstream'' (in a rotational sense) of the bar's end, which we may be seeing at the end of the ridgeline: there, the most intense \tco\ emission lies about 1.5\,kpc downstream from the RSA.

\begin{figure*}				
	\hspace{-0mm}\includegraphics[angle=-90,width=178.7mm]{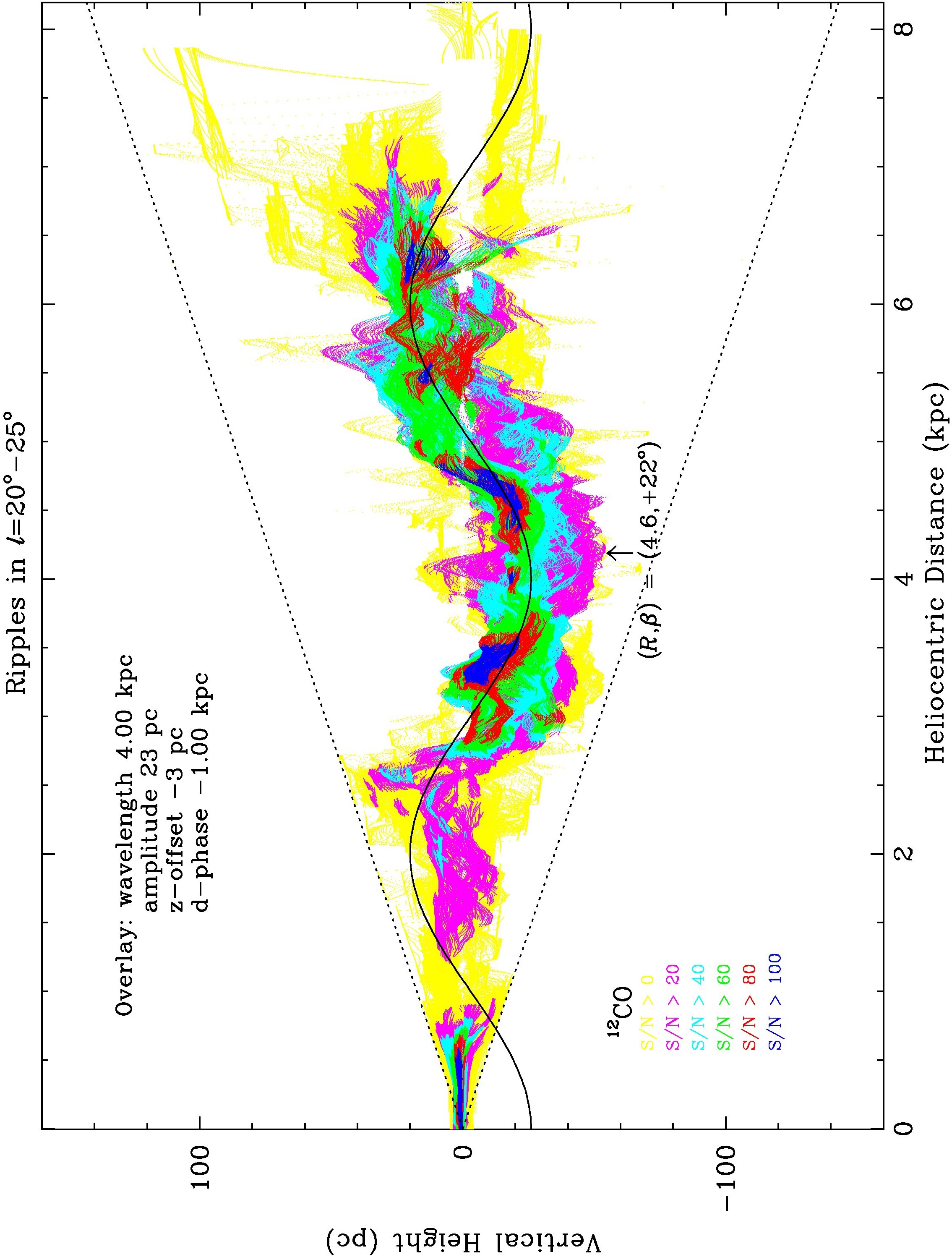}
	\caption{\footnotesize Mean projected height $\bar{z}$ (the first latitude moment) of FUGIN \tco\ data as a function of heliocentric distance, within a 
	5\degree\ sector of the 1Q \ld\ map.  Each coloured dot corresponds to 1 pixel in the \ld\ map, with the colour indicating the \tco\ brightness in S/N 
	units (the ``S/N $>$ 0'' label is an artifact of the plotting algorithm; it actually denotes S/N $>$ 5 in the SAMed data).  Overlaid as dotted black lines 
	are the latitude limits ($\pm$1\degree) of the data projected onto the physical $z$ scale, giving a vertical exaggeration for the plot of 20:1.  Also 
	shown are an approximate fit by eye (as labelled) of a sinusoid to the ridgeline of maximum \tco\ intensities, and the approximate Galactocentric 
	coordinates at the ``bottom'' of the ripple, which is $\sim$400\,pc outside the sharpest bend in the RSA at the end of the Bar as plotted in 
	Fig.\,\ref{xysmall}.
	\label{wiggles}	}
	\vspace{-0mm}
\end{figure*}

\subsection{Ripple Distribution and Latitude Structure}\label{midplane}

Also illuminating for this discussion is the first latitude-moment map in Figure \ref{xysmall} (middle panel) or {\color{red}Figure \ref{ftn-xy1}},
\begin{equation}
	\bar{z} = d \bar{b} = d \frac{\int I({\rm ^{12}CO})~b\,{\rm d}b}{I_{\rm CO,b}}~~~,
\end{equation}
where $b$ is measured in radians and we use the small-angle formula to convert the $b$ coordinate at each voxel to a linear height $z$ (since we have solved for $d$).  Here one sees the very clear banding of mean heights $\bar{z}$ across the inner Galaxy.  Proceeding outwards from the Sun up to a distance $d$ $\sim$ 2\,kpc, most molecular clouds have a mean height near 0\,pc.  Then, the mean height prominently dips across $d$ $\sim$ 2.5--4\,kpc to an extreme of $z$ $\approx$ --30\,pc below the midplane.  From $d$ $\sim$ 4--6\,kpc this reverses to a widespread rise up to $z$ $\approx$ +40\,pc, especially in the 4Q.  Beyond $d$ $\sim$ 6 or 7\,kpc, the pattern becomes less distinct as the detectable population thins, although in the 4Q, a hint of another dip at $d$ $\sim$7--8\,kpc and another rise at $d$ \gapp\ 8\,kpc can be discerned.  However, this isn't meant to imply that the $\bar{z}$ pattern is centred on the Sun: there are clear asymmetries to the above description, and the pattern is obviously more compressed in $d$ in the 4Q (i.e., the wavelengths are shorter) with larger amplitudes, and more stretched out (longer wavelengths) with smaller amplitudes in the 1Q.  

This pattern is amplified in the CfA data as seen in {\color{red}Figure \ref{cfa-xy1}}, where the antisymmetric pattern of ripples between the 1Q and 4Q, plus the changes in amplitude and wavelength (see below), are even plainer.  Taken together, these maps seem to provide robust confirming evidence for systematic deviations in the molecular layer from the mathematical midplane of $b$=0 \citep[called ripples by][also hereafter for convenience]{b25}.  This ripple pattern in $\bar{z}$ is even more obvious in the \ld\ projections than in \xy, but is slightly easier to interpret in the \xy\ maps, at least where overall Galactic cartography is concerned.

\begin{figure*}				
	\hspace{-0.2mm}\includegraphics[angle=-90,width=179mm]{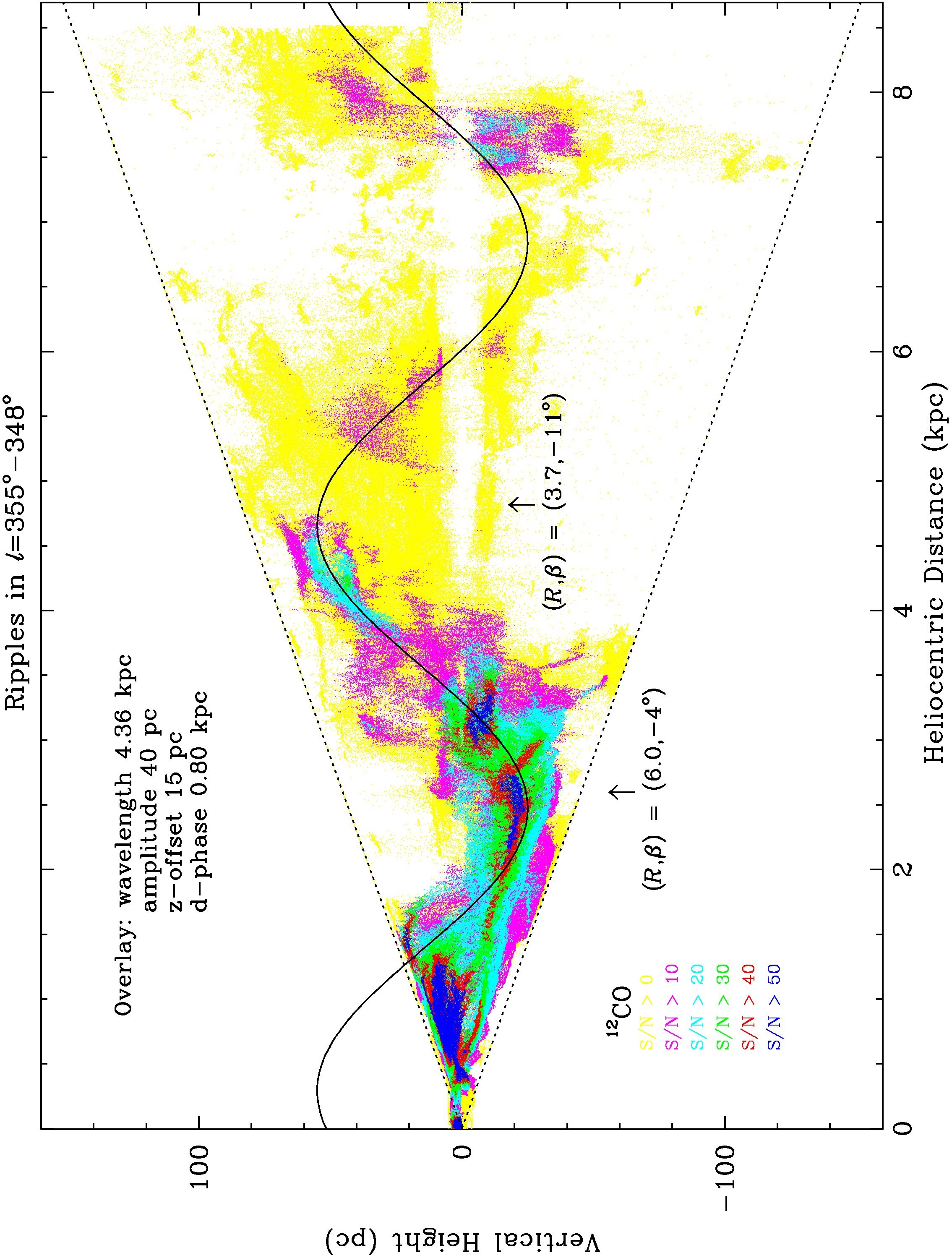}
	\caption{\footnotesize As in Fig.\,\ref{wiggles} but for $\bar{z}$ within a 7\degree\ sector of the 4Q \ld\ map of the \tco\ data from ThrUMMS 
	\citep[re-done from][their Fig.\,C31, due to the revised kinematic model used here]{b25}.  The sinusoidal fit here is different than in 
	Fig.\,\ref{wiggles}, and the RSA passes through the first ``bottom'' of the ripple, but other features of the plot are the same.  
	\label{wiggles4q}	}
	\vspace{-0mm}
\end{figure*}

One can also project the 3D \lbd\ data along the $l$ axis, colour-coded by the \ld\ integrated intensity, effectively forming a kind of ($d$,$b$) or ($d$,$z$) diagram, as done by \cite{b25}.  Examples are shown in {\color{red}Figures \ref{wiggles} and \ref{wiggles4q}} for narrow spans of longitude in the 1Q and 4Q, respectively. 
In the 1Q example, the fitted amplitude of the apparent wave or ripple is 23\,pc, about half the 40\,pc seen in the 4Q example; other longitude ranges in each Quadrant give similar results.  This dichotomy in the amplitudes between the two Quadrants can also be seen in either Figure \ref{xysmall} (middle panel) or Figures \ref{ftn-xy1} and \ref{cfa-xy1}, as reflected in the midplane displacements across the deprojected disk.

The pattern of the molecular layer's thickness, as measured by the second latitude-moment map $\sigma_{z}$, is shown in Figure \ref{xysmall} (bottom panel) or {\color{red}Figures \ref{ftn-xy2} and \ref{cfa-xy2}}.  In the FTN data, the pattern of thicknesses mimics that of \icob: there is some rough correlation between the two moments.  This suggests that on a ``large'' scale (say, $>$300\,pc), \tco\ emission counts the number of clouds within a given area rather the total molecular mass in that area.  
Otherwise, the molecular layer's thickness has a modal value around 20\,pc in the FTN data, although it is $\sim$20\% larger than this in the 1Q, and $\sim$20\% smaller in the 4Q.  The modal CfA thickness is about 38\,pc, presumably due to the lower spatial resolution in the CfA survey (10\,pc at 4\,kpc).  It is also seemingly more uniform across the disk, except for a $\sim$3$\times$1\,kpc area downstream from the end of the Bar, where $\sigma_{z}$ puffs up to surprisingly large values \gapp100\,pc.  Apart from this one area, these values are much thinner than the more classical value of 90\,pc derived by \cite{dht01}, but consistent with \cite{r19}'s value of 19\,pc for young OB stars, presumably due to the CfA data averaging over all the ripples in the disk.

\begin{figure*}
	\hspace{-0mm}\includegraphics[angle=-90,width=179mm]{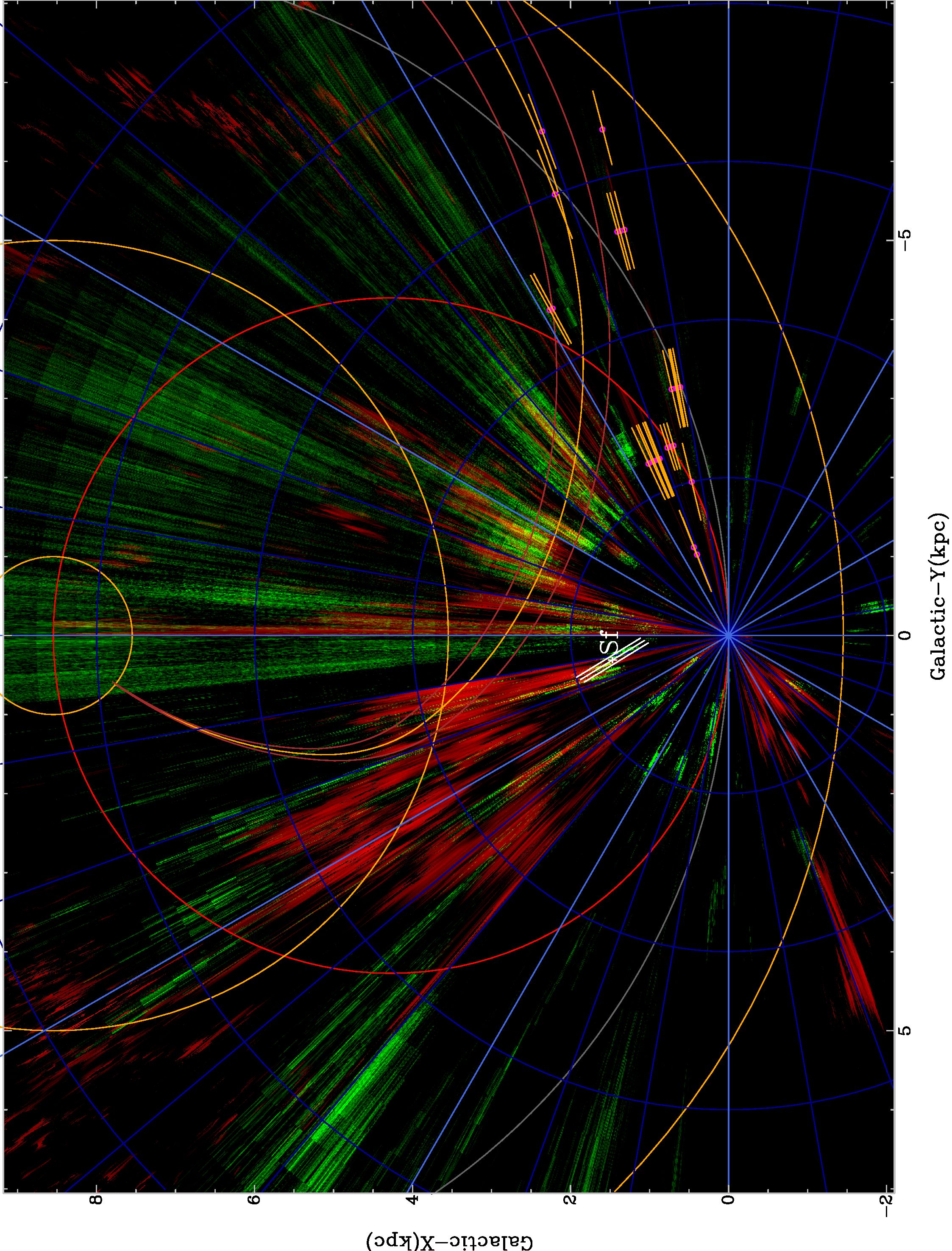}
	\caption{\footnotesize As in Fig.\,\ref{xysmall} (top panel) but now as a colour composite, with log(\icob) rendered in the red channel of the image, 
		and the composite Bayestar/DECaPS dust cube \citep{g19,z25}, effectively also on a log scale with units mag\,kpc$^{-1}$, rendered in green 
		(there is no blue image).  The dust cube was first truncated in latitude to match both ThrUMMS' and FUGIN's coverage, then the 0th moment 
		computed.  In this figure, the \citet{r19} spiral arms have been omitted for clarity, and galactocentric distance contours at 1, 5, and 10\,kpc 
		(orange) added.
	}\label{smalldust}
\end{figure*}

How are these ripples (Fig.\,\ref{ftn-xy1}) correlated with the zeroth-moment distribution (Fig.\,\ref{wftn-xy0}) and/or the spiral arms?  Comparing the two diagrams and starting around $\beta$ = 25\degree\ or $d$ = 5\,kpc near the end of the Bar in the 1Q, one sees that the ridgeline of maximum \icob\ is near an area of low $\bar{z}$ $<$ 0.  Continuing along the arc of the RSA as drawn in these figures, and also as the ridgeline approaches the \sct\ Arm, $\bar{z}$ remains below or near zero along this arc.  This is also true when looking ``downstream'' beyond the end of the Bar.  In contrast, $\bar{z}$ away from the RSA ridgeline rises to higher values $>$ 0.  Crossing into the 4Q data, the areas of highest \icob\ (initially tracking the \sct\ Arm until $l$ $\sim$ 330\degree) keep their somewhat neutral-to-negative $\bar{z}$, but points away from the ridgeline tend to retain a more positive $\bar{z}$.  Where $l$ $<$ 330\degree, however, the RSA's $\bar{z}$ trends to positive values, continuing to the edge of the ThrUMMS data at $l$ = 300\degree.  
In the the NANTEN+CHaMP sectors ($l$ $<$ 300\degree), the excursions to high or low $\bar{z}$ become more extreme in either direction, before the data become too sparse to define any obvious trends.  Nevertheless, one forms the impression of the RSA ridgeline appearing ``high'' in the 4Q but ``low'' in the 1Q.  

This pattern is also repeated rather precisely in the CfA data (Appendix \ref{cfaviews}), despite the latitude-vignetting of the higher-resolution surveys.  This again suggests that, as far as the overall structure of the molecular ISM is concerned, the FTN surveys have not missed very much.

\section{Comparison with the Dust Distribution}\label{dust}
\subsection{Where the Mass Distributions are Correlated}

Also interesting is the overall correspondence between the deprojected \icob\ map in Figure \ref{wftn-xy0} and the deep 3D deprojected dust extinction map of \citet[][their Fig.\,9]{z25}.\footnote{Based on a combination of \citet{z25}'s DECaPS (4Q) data and \citet{g19}'s Bayestar (1Q) data.  The DECaPS data, based on the DECaPS, VVV, 2MASS, and WISE projects, are deeper in both the optical and near-IR than the Bayestar data, based on Pan-STARRS1 and 2MASS.}  There is an especially good correspondence across much the the ThrUMMS area, but also for the Sgr feather in the 1Q, and at a lower overall mass surface density level, the CHaMP Regions.  Like \cite{b25}, \cite{z25} also noted the rather poor fit of their cloud distribution to any smooth log-spiral model.  Here we argue that the correspondence of both gas and dust tracers to a single, dominant, high pitch angle structure (i.e., via completely different methods) supports the idea that a single --- but ragged --- spiral arm appears to dominate the inner Galaxy's dense ISM structure.\footnote{Note that \cite{z25}'s projection of the \cite{r19} spiral arms onto their dust map does not align the \sct\ Arm with the RSA-ridgeline as here, due to their figure leaving unchanged the \cite{r19} $R_{0}$ scale.  The rescaling of $R_{0}$ to the VERA value salvages (somewhat) a 5\% larger version of \cite{r19}'s \sct\ Arm as an alternative to the RSA, but with a very different pitch angle (17\degree\ vs.\ 32\degree\ where they overlap).}

\begin{figure*}
	\includegraphics[angle=0,width=85mm]{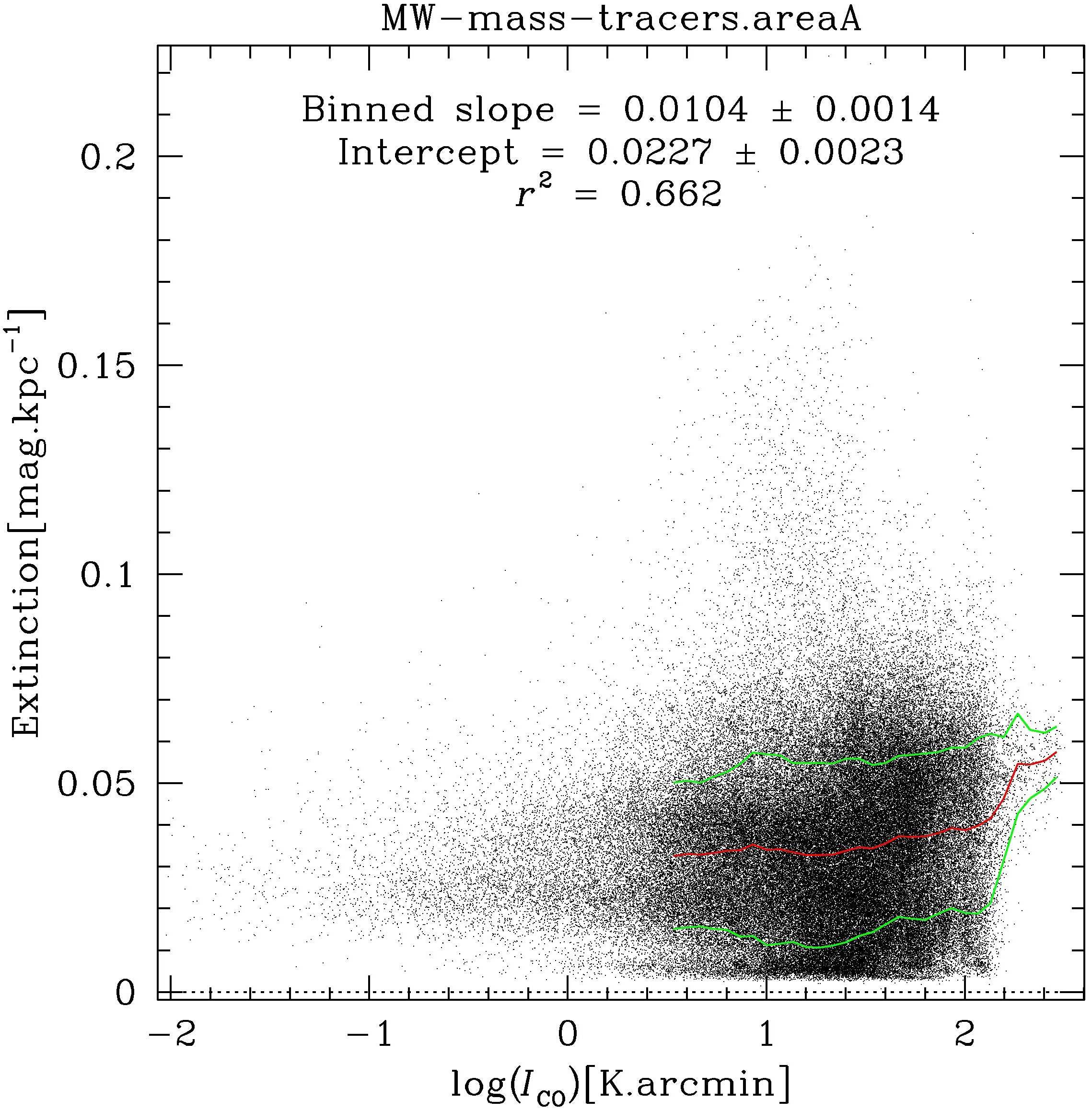}\hspace{9mm}  \includegraphics[angle=0,width=85mm]{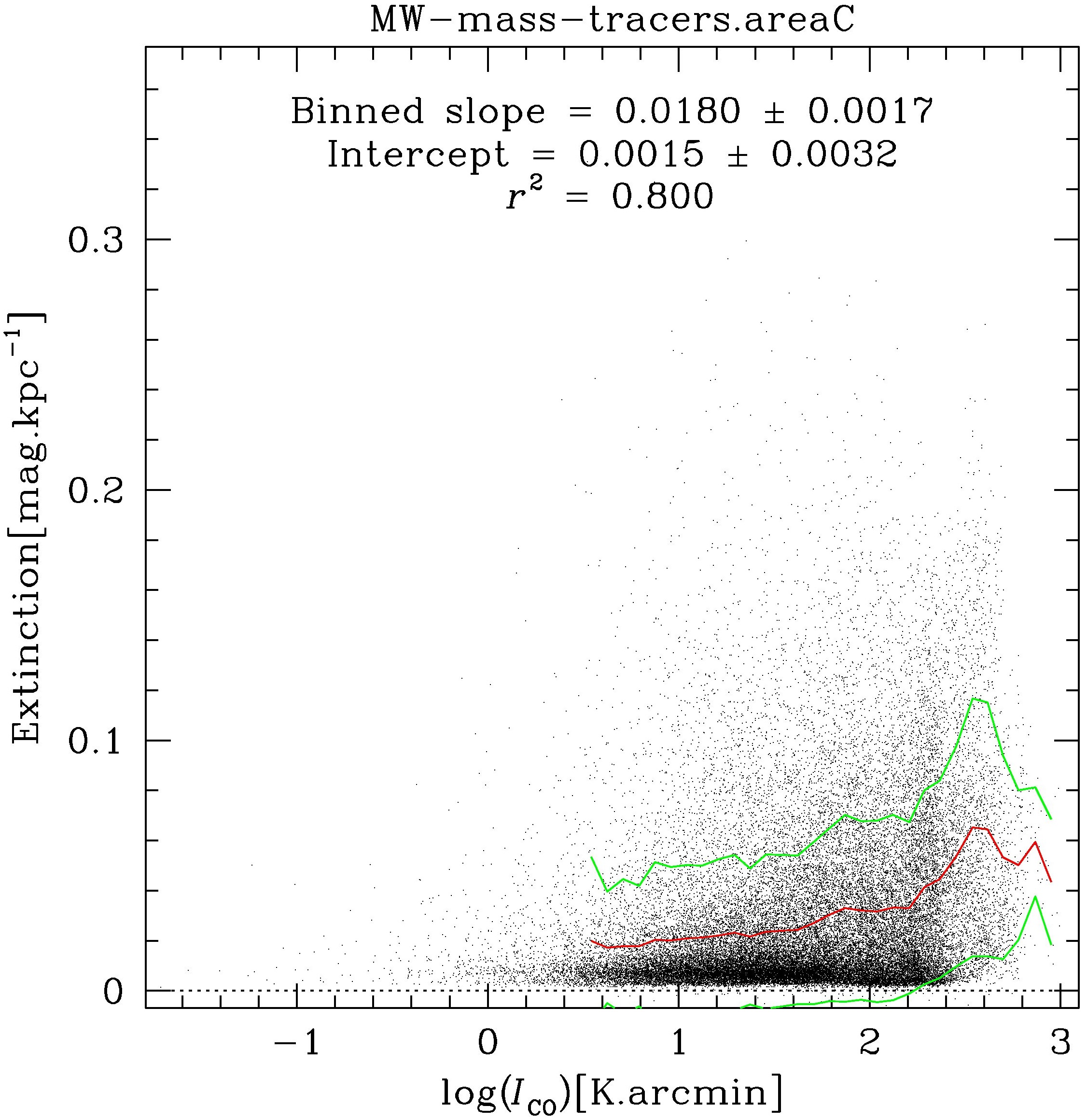}

	\vspace{-4mm}\hspace{-0mm}($a$)								\hspace{90mm}($c$)	

	\includegraphics[angle=0,width=85mm]{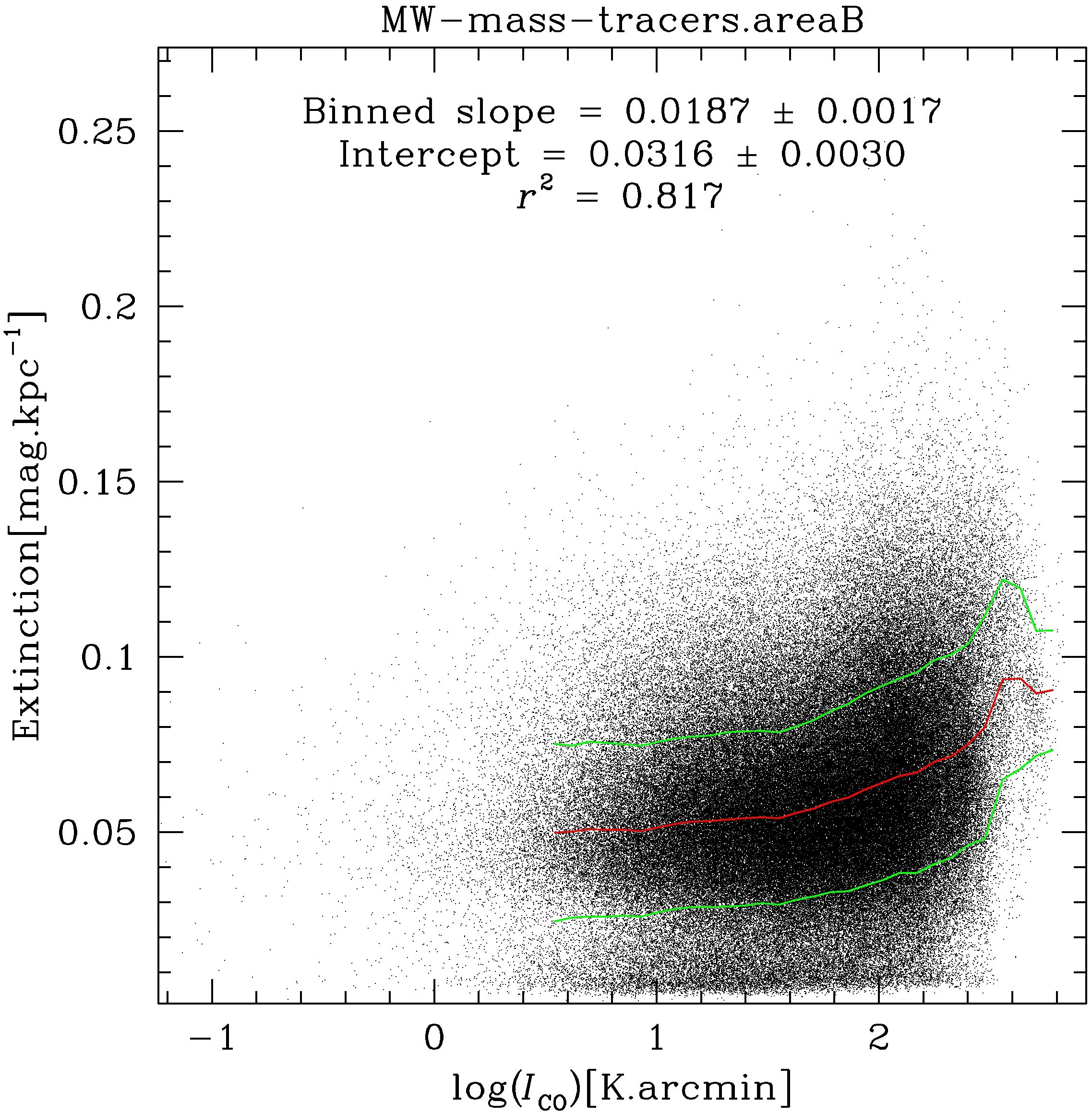}\hspace{9mm}  \includegraphics[angle=0,width=85mm]{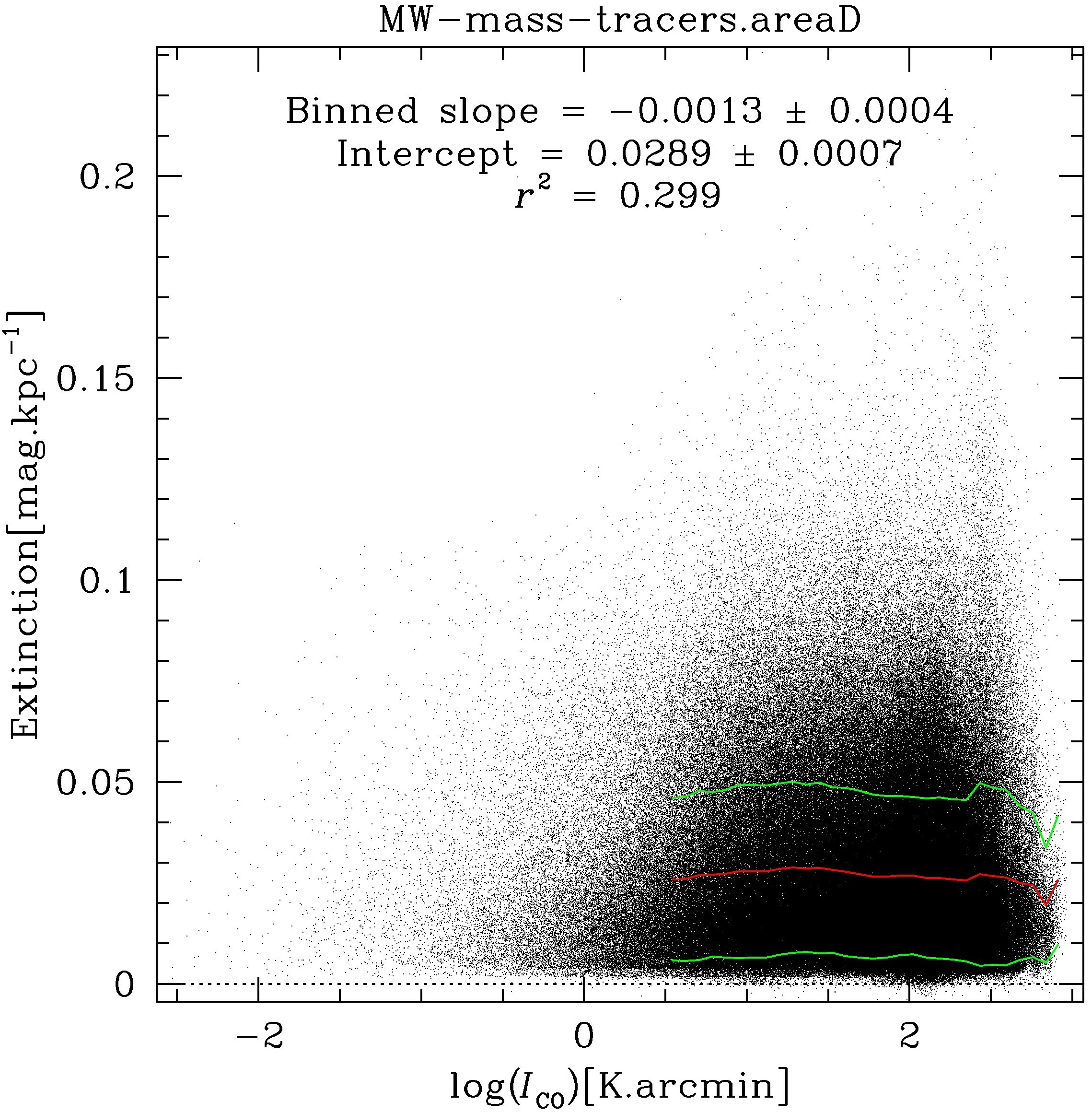}

	\vspace{-4mm}\hspace{-0mm}($b$)								\hspace{90mm}($d$)

	\vspace{0mm}
	\caption{\footnotesize Correlation of mass distributions based on \icob\ emission and dust extinction, for four sample areas across 
		the \xy\ plane of the Milky Way (panels $a$--$d$ for areas A--D as given in the text).  To make the data trends in each panel clearer, 
		the points were binned into 30 equally-spaced intervals between log(\icob) = 0.5 and 2.75, and the bins' mean ($\pm$1$\sigma$) 
		extinction values connected by red (green) line segments.  The labelled linear regression fits were then computed for the bins' 
		mean values.  Data/bins with log(\icob) $<$ 0.5 were not fit since they are uncorrelated (slope $\sim$ 0) everywhere.  Fits to the 
		unbinned data are similar except that the correlation coefficients are lower, $r^{2}$ = 0.01, 0.07, 0.07, 0.00 in areas A--D 
		respectively, due to the large intrinsic scatter in the extinction data.
	\label{masscorrel}}
\end{figure*}

A direct comparison of the full \xy\ maps as a two-colour composite is shown in {\color{red}Figure \ref{wftn-dust}}, with an inner Galaxy view in {\color{red}Figure \ref{smalldust}}.  In the 4Q where the two mass tracers coincide closely, the red (\tco) and green (dust) structures align well to give significant areas with a yellow colouring, e.g., from $l$ = 305\degree--345\degree\ at $d$ $\sim$ 3\,kpc, along the maximum-mass ridgeline traced by the RSA model.  To illustrate how good the structural correspondence is in the 4Q, {\color{red}Figure \ref{masscorrel}} shows the detailed point-to-point correlation within four sample areas across the \xy\ plane, from west to east as follows.

Area A: Longitudes 305\degree\ $>$ $l$ $>$ 280\degree\ and distances 1\,kpc $<$ $d$ $<$ 10\,kpc, covering the Carina Arm via CHaMP, NANTEN, ThrUMMS, and DECaPS data.

Area B: Longitudes 345\degree\ $>$ $l$ $>$ 305\degree\ and distances 1\,kpc $<$ $d$ $<$ 5\,kpc, covering much of the 4Q via ThrUMMS and DECaPS.

Area C: Longitudes 50\degree\ $>$ $l$ $>$ 10\degree\ and distances 0\,kpc $<$ $d$ $<$ 2\,kpc, covering the inner 1Q via FUGIN and Bayestar data.

Area D: Longitudes 50\degree\ $>$ $l$ $>$ 10\degree\ and distances 2\,kpc $<$ $d$ $<$ 10\,kpc, covering the outer 1Q via FUGIN and Bayestar.

Panels ($a$)--($c$) of Figure \ref{masscorrel} (areas A--C) all show fairly good correlations (albeit with a large scatter), confirming the visual impression of the 4Q in Figure \ref{smalldust}.  The slopes of the correlation between the gas and dust are highly significant at 7$\sigma$, 11$\sigma$, and 11$\sigma$ in each panel respectively, with correlation coefficients $r^{2}$ $\sim$ 0.7--0.8.  
Indeed, in all 3 cases, the correlation does not really take hold until log(\ico) \gapp\ 1.5, with the trends being essentially flat below this level.  Thus the true trends in higher column density clouds are even stronger than these statistics would suggest.

Visually, the area A correlation is less obvious in Figure \ref{smalldust} than that of area B, but that is partly due to the contrast settings in the rendering of the red/green image overlay.  Area C, which covers the inner 1Q and particularly the Sgr feather of \cite{k21}, also shows a good gas-dust correlation, 
statistically very similar to area B, although the extinction distribution is much more compressed to low values than in areas A or B.

\begin{figure*}
	\includegraphics[angle=-90,width=179mm]{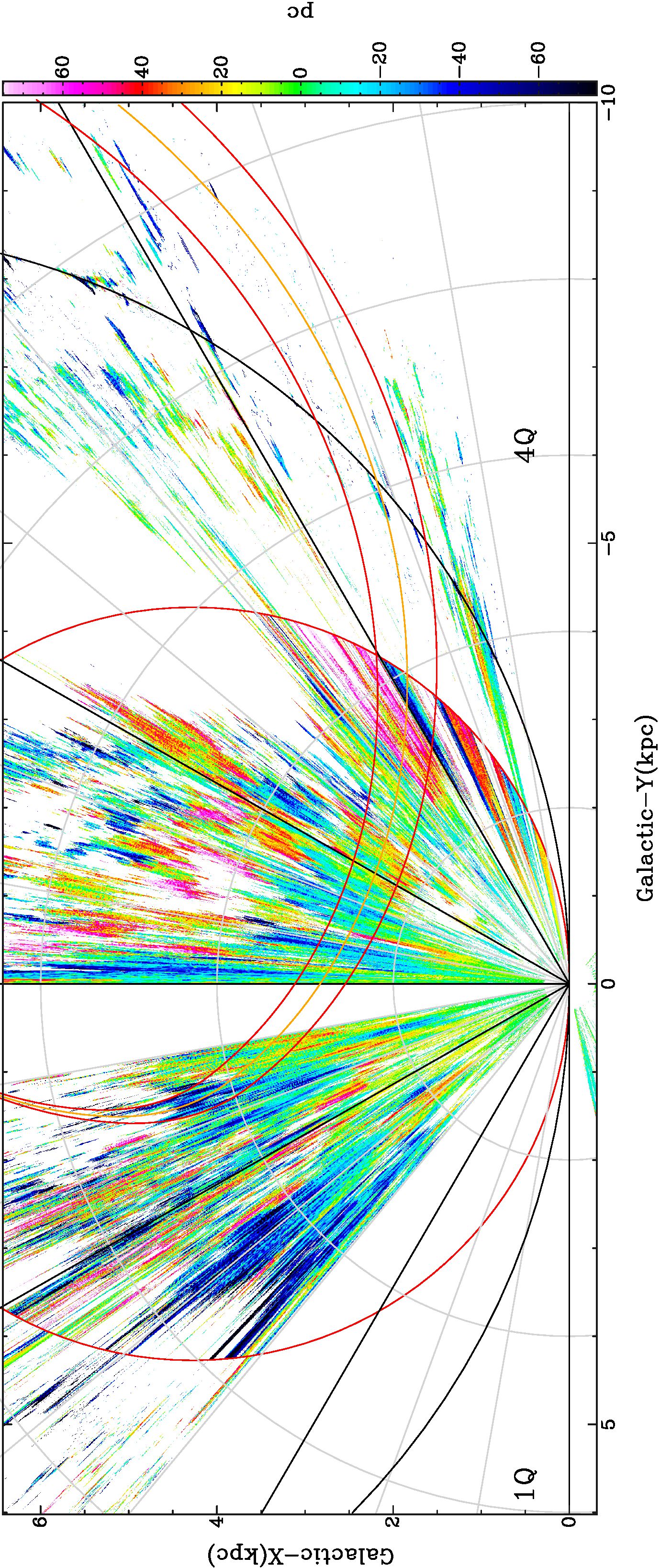}
	\caption{\footnotesize Similar map to Fig.\,\ref{xysmall}$b$, but now showing the difference in mean heights after subtracting the data 
		in Fig.\,\ref{dust-xy1} from the data in Fig.\,\ref{xysmall}$b$, yielding ($\bar{z}_{\rm gas}-\bar{z}_{\rm dust}$).  For simplicity, the 
		\citet{r19} spiral arms have also been omitted here.
	}\label{zdiff}
\end{figure*}

\subsection{Where the Mass Distributions are NOT Correlated}

As good as the correlations in the 4Q and inner 1Q appear to be, equally striking is the poor correlation between the gas and dust tracers in the outer 1Q.  Thus, panel ($d$) of Figure \ref{masscorrel} has a small $r^{2}$ = 0.30 and zero slope: the correlation has essentially collapsed!  This disparity between the impressive gas+dust correlation elsewhere and its lack in the outer 1Q is, on its face, very puzzling: the former implies we have essentially ``solved'' for the ISM's 3D structure in the 4Q+inner1Q, while the outer 1Q seems to suggest that either the dust extinction model or the rotation model (or both) are somehow ``wrong.''  However, there is a very reasonable explanation for this difference: the photometry upon which the Bayestar extinction results are based is not as deep as in the DECaPS data, meaning that the Bayestar dust map has rather large uncertainties beyond $\sim$2\,kpc \citep{z25}.  Thus, one might expect that if the extinction data in the 1Q were improved to the depth of that in the 4Q, one would recover the good gas-dust correlation seen in areas A--C.

A consequence of this large-scale \tco/dust correspondence is how it supports the kinematic model described in \S\ref{newmodel}.  
This suggests that to a large degree, the BVTFg model does ``solve'' the kinematic deprojection problem better than prior models, and might be useful in future studies of the inner Galaxy.

\subsection{Disparity between Mean Height Maps}

With the dust data cube, one can also compute higher latitude- or height-moments as was done with the molecular line data.  An example is given in {\color{red}Figure \ref{dust-xy1}}, which shows the dust extinction's first latitude-moment or mean height, $\bar{z}_{\rm dust}$.  Since this figure is on the same scale ($\pm$75\,pc) as other mean height maps, it is immediately obvious that the dust distribution is much flatter than the gas.  That is, while $z$-ripples still exist in the dust, they are much more subtle than those in the molecular gas.  This is made clear when one computes the difference ($\bar{z}_{\rm gas}-\bar{z}_{\rm dust}$) between the two mean-height maps, as shown in {\color{red}Figure \ref{zdiff}}.  This map looks generally quite similar to the $\bar{z}_{\rm gas}$ map alone (Fig.\,\ref{xysmall}, middle panel), indicating that the molecular gas ripples dominate over any such in the dust.

Upon closer comparison of Figures \ref{xysmall} and \ref{dust-xy1}, however, one sees that the (subtler) dust ripples often have opposite $z$ to the molecular gas ripples.  Superficially, this is disconcerting: if both trace the ISM mass distribution at some level, as is suggested by the moment 0 maps' correspondence at least in the 4Q (Fig.\,\ref{smalldust}), then the mean height distributions should at least be correlated.  However, this may be a small side effect of the reason for $\bar{z}_{\rm dust}$ being relatively flat, which is the more significant difference.

This difference probably arises in the different contributions to the total mass column from molecular and atomic gas.  That is, in the disk overall, and probably also at most specific locations \xy\ in the disk, the HI mass column exceeds the \htwo\ mass column by some factor of $\sim$a few.  Supposing further that the gas and dust are reasonably well-mixed, so that for example, the gas-to-dust ratio $\gamma$ is roughly constant, one would expect the $z$-distribution of the dust to match that of the dominant HI distribution more closely than that of the lesser \htwo\ distribution.  Since the HI layer is spread more widely and smoothly than the molecular layer (scale height typically $\sim$100\,pc), it should not be surprising that $\bar{z}_{\rm dust}$ $\sim$10\,pc is relatively closer to 0 than the more clumpy molecular layer's $\bar{z}_{\rm gas}$ $\sim$30\,pc.

If this is the correct picture, the additional detail that the dust ripples seem slightly opposite in sign to the molecular ripples may be easily explained.  That is, the dust extinction is measured against stars that are visible in the optical and near-IR.  Most molecular clouds mapped by the CODEX surveys are typically too opaque to allow such stars to be visible behind them, at least in their denser portions where \ttco\ and even \ceto\ are detected in addition to the \tco.  Thus, the dust extinction data will have a slight bias against tracing such clouds, compared to the more diffuse HI clouds and to any general extinction in the ISM away from most clouds, such as in patches of hot coronal gas.

That this is actually the case can be seen in Appendix A of \cite{z25}, their Figure 15e.  They describe how the denser interiors of clouds are so opaque that there is a deficit of extinction data (i.e., often none) covering their projections on the sky: they appear as ``shadows'' in the extinction maps.  Therefore it is fair to say that the extinction data are biased away from dense cloud interiors.  Statistically, what this means is that the distribution of the extinction (especially as a function of $z$) will anti-correlate with that of the denser gas, on the cloud ($\sim$10\,pc) scale.

While this picture likely explains the somewhat {\em different} $z$ distributions and embedded ripples of the dust and gas, it does not contradict the overall mass {\em correlation} of the two tracers across the disk (Figs.\,\ref{smalldust}, \ref{masscorrel}), since there we are averaging over kpc$^{2}$ areas much larger than the $\bar{z}$ scales of the disk's thickness.  At this coarser level when averaging over smaller ($<$100\,pc) structures, mass appears to scale with mass.
 
\section{Conclusions}\label{concl}

This paper presents a new meta-survey of the disk of the Milky Way, the CODEX, which is planned to combine and harmonise the data from several multi-species molecular line surveys from the last $\sim$15 years.  This will put the (2D or 3D) data from all included surveys onto a consistent grid, in order to address a number of whole-Galaxy science questions.

The first version of the CODEX combines information from 5 prior surveys, NANTEN, CHaMP, ThrUMMS, FUGIN, and MWISP, focusing mainly on the inner Galaxy (1+4Q) and using these to refine models of Galactic rotation.  The values for $R_{0}$ and $\Theta_{0}$ are optimised over all the survey data, and 
a modification to the kinematic distance formulae for nearby clouds (i.e., $d$ $<$ 1\,kpc) is developed to better match these clouds' true distances, despite their non-rotational motions.  Future studies could extend this idea to any number of ``Regional Standards of Rest'' (RSRs) that have their own peculiar (non-circular) velocities, such as a spiral arm with local streaming motions, an expanding SFR/GMC complex, etc.

Based on this updated rotation model, the cloud distributions across the inner Galaxy are examined.  The results are as follows.

\textbullet\ The 1Q (FUGIN) data are shown to have a consistent overall spiral mass distribution, and very similar pattern of ripples in the molecular layer (amplitude $\sim$30\,pc, wavelength $\sim$4\,kpc), to the recent 4Q (ThrUMMS) results of \citet{b25}.

\textbullet\ A new, high pitch angle, Ragged Spiral Arm (RSA) model is defined based on the inner Galaxy molecular line data, overlapping only $\sim$1\,kpc of the \citet{r19} definition of the Sct-Cen Arm, and only when that model is rescaled to the larger $R_{0}$ used here.

The gas distributions in mass and height are also compared with those from the dust extinction maps of \citet{g19} and \citet{z25}, and we find the following.

\textbullet\ The higher-sensitivity DECaPS extinction data are well correlated on $\sim$kpc or larger scales in \xy\ where they overlap with the 4Q molecular data; the less-deep Bayestar extinction data are also well correlated with overlapping 1Q molecular data for distances up to $\sim$2\,kpc.  Beyond this distance the Bayestar/FUGIN correlation is poor, apparently due to sensitivity limits in the extinction data.

\textbullet\ There is little evidence for a continuous Sgr-Car Arm in either the molecular line or dust extinction data, rather there seem to be several discontinuous features such as the Sgr feather of \citet{k21} and the various Carina complexes of NANTEN and CHaMP \citep{b11}.

\textbullet\ The dust and gas distributions in mean height $\bar{z}$ show smaller-scale mismatches than the large-scale correlation in \xy, such that the ripples as traced by dust  (amplitudes $\sim$10\,pc) are more muted than, and opposite in sign to, those seen in the gas (amplitudes $\sim$20--40\,pc).  This can be understood as the dust extinction tracing well the total (atomic+molecular) diffuse gas distribution, but not tracing the dense molecular gas distribution so well; the molecular line data then have the opposite bias in this regard.  If this interpretation is correct, the gas and dust height distributions can also be considered mutually consistent.


\begin{acknowledgments}
This work would not have been possible without the commitment to publicly available data by the PIs and contributors of the various surveys, and we thank and commend them for the significant effort in creating the data products and archives.  PJB gratefully acknowledges support from NSF grant AST-2206584, and warmly thanks Mark Wieringa for his outstanding support of Miriad, Bob Benjamin for inspiring the latitude/height analysis which led to the discovery of  the ripples, Catherine Zucker for help accessing and interpreting the 3D dust data, and Ana Duarte Cabral \& Paul Goldsmith for helpful comments on the manuscript.  
\end{acknowledgments}

\vspace{2mm}{\em Facilities:} {Mopra (MOPS), NANTEN, Nobeyama Radio Observatory 45m, Purple Mountain Observatory 13m,
Center for Astrophysics 1.3m}

\vspace{2mm}{\em Software:} {karma \citep{g97}, Miriad \citep{s95}, SuperMongo \citep{lm00} }



\appendix

\section{Kinematic Deprojections and Updates to Rotation Models}\label{rotmodels}
\subsection{Recent Work}

This rotation parameter analysis is similar to the work of \cite{b25} with the ThrUMMS data alone, except that there, the longitude range was limited to 300\degree--360\degree\ in the Fourth Quadrant (hereafter 4Q).  This is now extended to $l$ = 280\degree--300\degree\ in the 4Q \citep[][and references therein]{y05}; to $l$ = 10\degree--50\degree\ in the 1Q \citep{um17}; and to $l$ = 105\degree--150\degree\ in the 2Q \citep[][and references therein]{s20}.

\cite{b25} showed that, across the ThrUMMS $l$ range, none of the existing rotation models or scales gave a very satisfactory fit to features of the \lv\ pattern as observed in the 4Q.  They started with the 1970s-era IAU definitions of the LSR (which still sets the \vlsr\ scale today) based on the understanding of the Sun's peculiar motion (\uvw) at that time, plus the IAU-standard scale factors $R_{0}$ and $\Theta_{0}$.  They also considered modifications to these parameters by the BeSSeL project \citep{r19}, by {\em Gaia} \citep{bob23}, and by VERA \citep{o24} (their Table C1).  A subset of these results is also presented here ({\color{red}Table \ref{rotpars}}) for comparison with the adjustments below.

\begin{figure}
	\hspace{40mm}\includegraphics[angle=180,width=90mm]{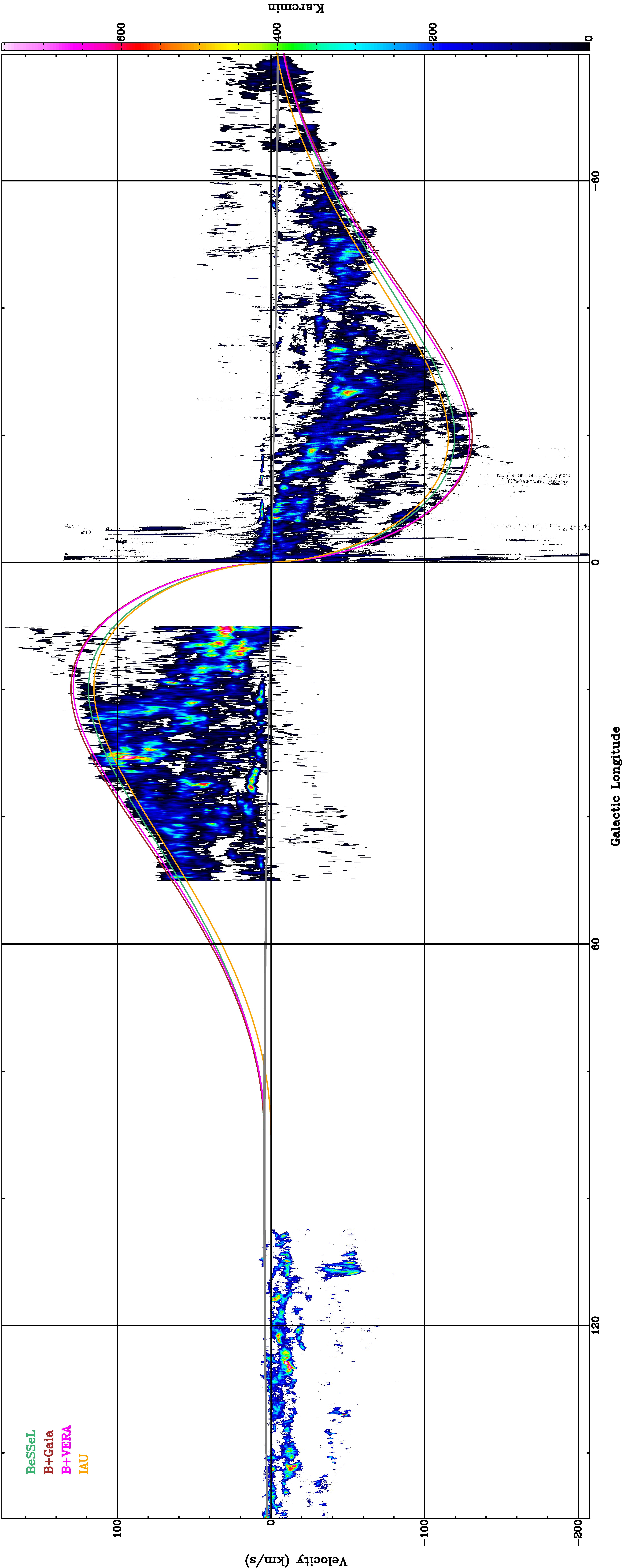} 

	\vspace{-1mm}
	\caption{\footnotesize Composite \lv\ diagram of latitude-integrated \tco\ emission in the MWISP, FUGIN, ThrUMMS, and NANTEN surveys.  
	Overlaid as an orange curve is the tangent velocity with the original IAU value for $\Theta_0$ = 247\kms; the $l$ axis is the IAU heliocentric 
	solution for \vlsr\ = 0 by definition.  The other coloured tangent curves are for the labelled rotation models (see Table \ref{rotpars}), with 
	their almost identical \vlsr\ = 0 curves in grey.  We encourage examining these high-resolution images at high magnification.
    \label{oldLVmodels}}
\end{figure}

The mismatch between the IAU parameters and the \lv\ data is readily apparent in {\color{red}Figure \ref{oldLVmodels}}.  First, the tangent curve (orange), which by definition should be perfectly antisymmetric in both the 1Q and 4Q, falls well short of the extreme velocity envelope in both inner Quadrants.  This problem is most severe in 310\degree\ $>$ $l$ $>$ 280\degree\ (the Carina tangency).  Although this could be remedied somewhat by simply scaling up $\Theta_0$ by $\sim$10\%, the ThrUMMS data further reveal a very distinct, narrow-line population of clouds (hereafter NLCs) which are local to the Sun (i.e., heliocentric distances \lapp 300\,pc and typical $|$\vlsr$|$ $<$ 20\,\kms).  As such, \cite{b25} argued that they should {\em define} \vlsr\ = 0, since they likely indicate a sinusoidal offset from the IAU \vlsr\ definition, equivalent to a revised set of (\uvw) parameters for solar motion (SM).  Provided that similar clouds could be found in other Quadrants consistent with the same sinusoid, they argued that such clouds would establish a truer marker of \vlsr\ as the solar neighbourhood's most extreme Pop I sample.  However, adjusting the (\uv) solutions in this way has the additional effect of making the tangent velocity envelope asymmetrical between the 1Q and 4Q.  While the \vlsr\ = 0 fit is improved, the IAU value for $\Theta_0$ is even more untenable than before.  (This is discussed further in \S\ref{newmodel} below.)

Fortunately, the BeSSeL, {\em Gaia}, and VERA fits all suggested appropriately larger values than the IAU's $R_0$ and $\Theta_0$ scale parameters, and at least within the 4Q, \cite{b25} found a suitable combination of scale and rotation parameter adjustments to make their overall ``BGT'' model fits\footnote{The BGT model is not given here; see \cite{b25}'s Table C1 and Figures C1--C5 for full details.} quite reasonable for the ThrUMMS data. 

\begin{deluxetable}{lcccccc} 
\tabletypesize{\footnotesize}
\tablecaption{Kinematic parameters for Solar motion \& Galactic rotation}
\tablehead{
	\colhead{Kinematic} & \colhead{Galactocentric} & \colhead{Orbital} & & \multicolumn{3}{c}{Solar Motion\tablenotemark{$^{a}$}} \\
	\cline{5-7} 
	\colhead{Model*} & \colhead{Radius $R_0$} & \colhead{Speed $\Theta_0$} & & \colhead{$u_0$} & \colhead{$v_0$} & \colhead{$w_0$} \\
	 & \colhead{(kpc)} & \colhead{(\kms)} & & & \colhead{(\kms)} & 
}
\startdata
	IAU-LSR & 8.15\tablenotemark{$^{b}$} & 236.3\tablenotemark{$^{b}$} & & 10.27 & 15.32 & 7.74 \\
	B & 8.15 & 236.3\tablenotemark{$^{c}$} & & 10.6 & 10.7\tablenotemark{$^{c}$} & 7.6 \\
	BG & 8.24 & 257.3\tablenotemark{$^{d}$} & & 10.6 & 10.7\tablenotemark{$^{d}$} & 7.6 \\
	BV & 8.55\tablenotemark{$^{e}$} & 248.3\tablenotemark{$^{e}$} & & 10.6 & 10.7\tablenotemark{$^{e}$} & 7.6 \\
	BVTFg & 8.55 & 248.3 & & 3.6\tablenotemark{$^{f}$} & 10.7 & 7.6 
\enddata 
\tablenotetext{*}{The acronyms listed designate different rotation models based on combined results from the following projects: B = BeSSeL \citep{r19}; 
		G = {\em Gaia} \citep{bob23}; V = VERA \citep{o24}; T = ThrUMMS \citep{b25}; F = FUGIN \citep{um17}; g = this work, Eq.\,(A2).} 
\tablenotetext{a}{Additional motion relative to LSR, parallel to cartesian coordinates $x$,$y$,$z$.} 
\tablenotetext{b}{Assumed from BeSSeL; the IAU definition of LSR converts to the (\uvw) shown on this line.} 
\tablenotetext{c}{For BeSSeL, ($\Theta_0+v_0$) = 247\,\kms\ is more strongly constrained than either $\Theta_{0}$ or $v_{0}$ separately.} 
\tablenotetext{d}{For the Gaia Cepheids, ($\Theta_0+v_0$) = 268\,\kms\ is combined with the BeSSeL value for $v_0$.} 
\tablenotetext{e}{For VERA, $\Omega$\solar\ = $(\Theta_{0}+v_{0})/R_0$ = 30.30\,\kms\,kpc$^{-1}$ is {\bf far} more strongly constrained than the individual parameters.  Their separate solution for $R_0$ (as listed, $\pm$15\%) then gives ($\Theta_0+v_0$) = 259.0\,\kms, where $\Theta_0$ obtains assuming the BeSSeL value for $v_0$.} 
\tablenotetext{f}{As with Note e, but using an additional gaussian correction to $u_0$ for clouds near the Sun, Eq.\,(A2).} 
\vspace{-0mm}\label{rotpars}
\end{deluxetable}\vspace{0mm}

\vspace{-0mm}
\subsection{Towards An Improved Model}\label{newmodel}

This issue is now revisited with the advantage of a much wider lever arm in longitude, namely $\Delta$$l$ = 230\degree\ across three Quadrants, instead of only 60\degree\ in one Quadrant.  While a population of NLCs can easily and gratifyingly be seen in the 1Q data, they do not quite line up with the BGT \vlsr\ = 0 curve, with $\sim$half of them having ``forbidden'' velocities by $\sim$5\kms. 
This was apparently the result of overfitting the 4Q NLC data somewhat, giving a $v_{0}$ that was 5.0\kms\ lower than the BeSSeL solution, and a $u_{0}$ that was 8.5\kms\ lower.  So the first option was to follow \cite{b25}'s approach and adjust the BGT model, mainly by reference to the FUGIN NLCs near \vlsr\ = 0.  (An alternative approach is also discussed further below.)

Ignoring the tiny contribution of $w_0$ to this problem, the (\uv) parameters set the amplitude of the \vlsr-correcting sinusoid in \lv\ space, at $l$=0\degree\ and $l$=90\degree, respectively.  
Considering only these parameters, a new \vlsr=0 curve {\em should} ideally lie at an observed \vlsr\ just less than all of these clouds.  However, adjusting $u_0$ higher is strongly disfavoured, because that lowers the new \vlsr=0 curve (which moves in the opposite direction in an $lV$ diagram to the \uv\ adjustments) to below the 4Q NLCs as well, and they provide an even stronger constraint on $u_0$ than do the 1Q NLCs.

While $u_0$ could be raised somewhat, the resolution to this wider fit was to raise $v_0$ more substantially, bringing it much closer to the BeSSeL value.  This has the effect of flattening the asymmetry in the new \vlsr\ curve across $l$=0, and recovers a new \vlsr=0 solution where the degree to which any of the 1Q or 4Q NLCs have forbidden velocities is only $\sim$2\kms.

The 2Q data are less useful in this exercise, in the sense that the location of an NLC population in \lv\ space is a bit unclear.  However, keeping $v_0$ too small also placed any new \vlsr=0 curve a few \kms\ more deeply into the rather sharp positive-\vlsr-envelope of the 2Q clouds than seemed reasonable.  The outer Quadrants are also no help in defining the tangent velocities, since by definition, they have none.  So with the adjusted the (\uv) parameters as described above, optimising $\Theta_0$ with one of the BeSSeL, {\em Gaia}, or VERA values was the next option.  ($R_0$ has little impact in fitting data in the \lv\ diagram).  The result is that \cite{b25}'s slight preference for {\em Gaia}'s $\Theta_0$ over VERA's is now reversed.  So with the new (\uv) values as described above, the more preferred rotation model and parameter set is ``BVTF.''

However, this underscored a fundamental incompatibility between the assumption of the ``true'' \vlsr\ being traced by the local cloud population, which introduces a constant asymmetry in (say) the heliocentric distance curves that can be overplotted on an \lv\ diagram, and the close-to-antisymmetric tangent velocity envelope in the molecular line data.  In other words, almost {\em any} deviation from the IAU (\uvw) parameters will produce an equivalent change in the symmetry of the tangent curve, and ultimately, one cannot have it both ways.\footnote{The changes induced by the BeSSeL, Gaia, or VERA models, visible in the grey \vlsr\ = 0 curves of Fig.\,\ref{oldLVmodels}, are small enough to avoid this objection.}  The solution lies in recognising that the local clouds may not, in fact, represent a true \vlsr\ = 0, but might rather be part of a $\sim$10\kms\ amplitude local flow superimposed upon an LSR better defined by the stellar and maser data \citep[such as by][]{r19}.  However, this does not explain (for example) the much wider mismatch across kpc scales between the molecular clouds along the Carina tangency and any of the tangent curves in Figure \ref{oldLVmodels}.

Consider the ripples already reported by \cite{b25}, and at least qualitatively matched by corresponding results in other studies \citep{bin24,pkd25,a25}.  The emerging picture is that the most recent perigalactic passage of the Sgr dwarf has induced an $m$=1 disturbance in the otherwise flat disc of the Milky Way.  Then, objects in the Milky Way's disc currently have both a circular orbital motion, and oscillatory motions in all 3 dimensions around their own localised standard of rest (perhaps better called a Regional Standard of Rest, or RSR).  The RSR is likely defined to some extent (perhaps largely) by the $m$=1 disturbance itself; thus, the RSR in one part of the disc will not be the same as the RSR in another part, or as the Sun's LSR.  In fact, it may be that the amplitude of these local oscillations scales with location relative to the $m$=1 wave, and at some location(s) between the Sun and the Galactic Centre, the oscillations may be locally small.  If this were so, it could provide a potential solution to the above conundrum.  The discussion that follows is then aimed not so much at supplanting the definition of our LSR as \citet{b25} tried to do, but rather at a mechanism to incorporate local flows of the ISM into the general kinematic formalism, and allow derivation of kinematic distances even where there may be observable non-rotational motions (such as with the NLCs) that place clouds away from their own RSRs in an \lv\ diagram.

With {\em our} LSR, the current objective is then to derive non-rotational corrections that are assumed to apply to clouds only near the Sun.  Then, any such effect must diminish with heliocentric distance, for the purpose of deriving a ``true'' heliocentric distance $d$.  In other words, one only needs a correction for the local clouds' peculiar motions at small $d$.  This numerical correction to LSR at any general distance $d$ is not currently defined, but assuming it falls to 0 at some distance between us and the Galactic Centre (i.e., sufficiently beyond where we can see the NLCs), one can simply take the best local-clouds version of (\uvw), and scale it down with $d$.  Thus, near the Sun, the envelope of this LSR correction
is 1, but at $d$ $>$ 0
it is $<$1.  Further, because of the near-antisymmetry of the ensemble of tangent velocities, one might expect the envelope to approach 0 even at fairly small distances $d$, and certainly towards the Galactic Centre.  This very simple justification for a scaled \vlsr\ correction can be numerically incorporated into kinematic distance calculations as follows.

Let the functional form for the axisymmetric rotation velocity as a function of Galactocentric radius $r$ be $V_{\rm rot}(r)$; for example, a Universal Rotation Curve \citep{p96} as described by \cite{r19}.  Then for a cloud at a given longitude $l$ with a measured \vlsr\ (i.e., using the standard IAU-LSR),
\begin{equation}
	V_{\rm LSR}(l,d) = V_{\rm rot,proj}(r(l,d)) - [v_{\rm corr}{\rm sin}(l) + u_{\rm corr}{\rm cos}(l)] - s(l,d),		
\end{equation}
where $V_{\rm rot,proj}$ is the projected circular rotation velocity at longitude $l$ and radius $r$, and $r$ in turn is a function of $l$ and $d$, and the corrections from the IAU to the preferred LSR are given by $u_{\rm corr}$ and $v_{\rm corr}$, as described by \cite{r19}.  

Introduced in Eq.\,(1) is an additional correction function $s$, meant to be applied only to clouds local to the Sun (i.e., $d$ \lapp\ 1\,kpc, such as the NLCs mentioned above), and to otherwise leave the full kinematic solutions from prior works (such as BeSSeL) unmolested:
\begin{eqnarray}
	s(l,d)  &=&  e^{-(d/d_{s})^{2}} [v_{s}{\rm sin}(l) + u_{s}{\rm cos}(l)]~,		
\end{eqnarray}
where conservatively, $d_{s}$ = 0.5\,kpc (larger values are possible; more on this in \S\ref{zoom}).  The gaussian form ensures a tapered scaling to this additional correction, such that it essentially disappears beyond $\sim$2$d_{s}$ regardless of the values chosen for $u_{s}$ and $v_{s}$.  This scaled taper then operates mathematically in the same way as the \cite{r19} LSR corrections, but focused on only a small area in a way that is physically consistent with the overall BeSSeL \uvw\ solutions. 

This is a somewhat different approach than \cite{b25}, who varied $u_{\rm corr}$ and $v_{\rm corr}$ directly and applied these changes to all their kinematic distance calculations.  To some extent, that undoes the global calibrations that went into $u_{\rm corr}$ and $v_{\rm corr}$ in the first place.  Introducing the tapered offset $s$ as a separate adjustment to Eq.\,(A1) is a more self-consistent approach, when applied to the specific situation of non-rotational flows among nearby clouds.

With this alternative formulation, the task is to fit values for $u_{s}$ and $v_{s}$ based on the location of the NLC clouds in \lv\ space.  In Figure \ref{oldLVmodels}, the mismatch between the NLCs and the prior \vlsr\ solutions is most prominent in the 4Q, making the NLCs' velocities ``forbidden'' by (e.g.) $\sim$7\,\kms\ across $l$ $\approx$ 340\degree--354\degree, equivalent to a distance offset of $\sim$0.4\,pc.  This can be satisfactorily remedied by setting $u_{s}$ = --7\,\kms\ and leaving $v_{s}$ = 0.  Then, the NLCs in the 1Q also straddle the new \vlsr\ = 0 curve reasonably well. 

In reality, the true functional form of the scaling factor $s$ may well be more complicated than the simple form of Eq.\ (A2).  It could even be oscillatory itself, if it is to take into account potential 3D wave motions in the disc, or also possibly longitude-dependent, for example to take into account streaming motions (another kind of ``LSR'') along spiral arms, such as may be present in the Carina tangency.  But as long as $s$ can be described algebraically in some form, the kinematic distance can, in principle, still be computed, potentially with higher accuracy for local clouds than standard models with $s=0$.

\vspace{-0mm}
\section{Software}\label{code}

\vspace{-0mm}
The basic functionality of the various rotation models and their parameters as described in \S\ref{rotparsumm} has been incorporated into a SuperMongo \citep{lm00} program \texttt{lvfeat.sm}.  This is useful for making simple plots in the various Galactic coordinate systems, both as standalone encapsulated postscript (suffix .eps) files, and also as annotations files (suffix .ann) for overlays onto images rendered with {\em kvis} \citep[part of the {\em karma} package;][]{g97}.  Both products were used in many of the figures in this paper.  This SM program, first written by \citet{b25}, is fast and easy to use, and is made public herewith at \texttt{https://github.com/DrThip/lvfeat.git}.  Readers can find there a general description of what the program does; full documentation appears in the code itself.

In modern astronomical data processing, use of the Python language (via astropy, for example) has become very widespread, but a Python version of \texttt{lvfeat.sm} is not available.  SuperMongo syntax is relatively straightforward, however, so interested readers may wish to port \texttt{lvfeat.sm} to Python, and submit it as a new version of \texttt{lvfeat.sm} at the GitHub site above.

A more complex program \texttt{MWdist.csh} to actually transform the 2D and 3D data files between \lv\ and \ld\ or \xy, based on the formulae behind \texttt{lvfeat.sm} but managing the data regridding as well, was also written by \citet{b25}.  However, \texttt{MWdist.csh} requires serious computational power to perform the transformations.  It uses a combination of c-shell file management and \textsc{Miriad} \citep{s95} tasks for the data processing, and is available by collaboration.  
\vspace{0mm}

\clearpage
\section{Full-Scale CODEX Moment Maps}\label{wideview}

\begin{figure*}
	\includegraphics[angle=-90,width=179mm]{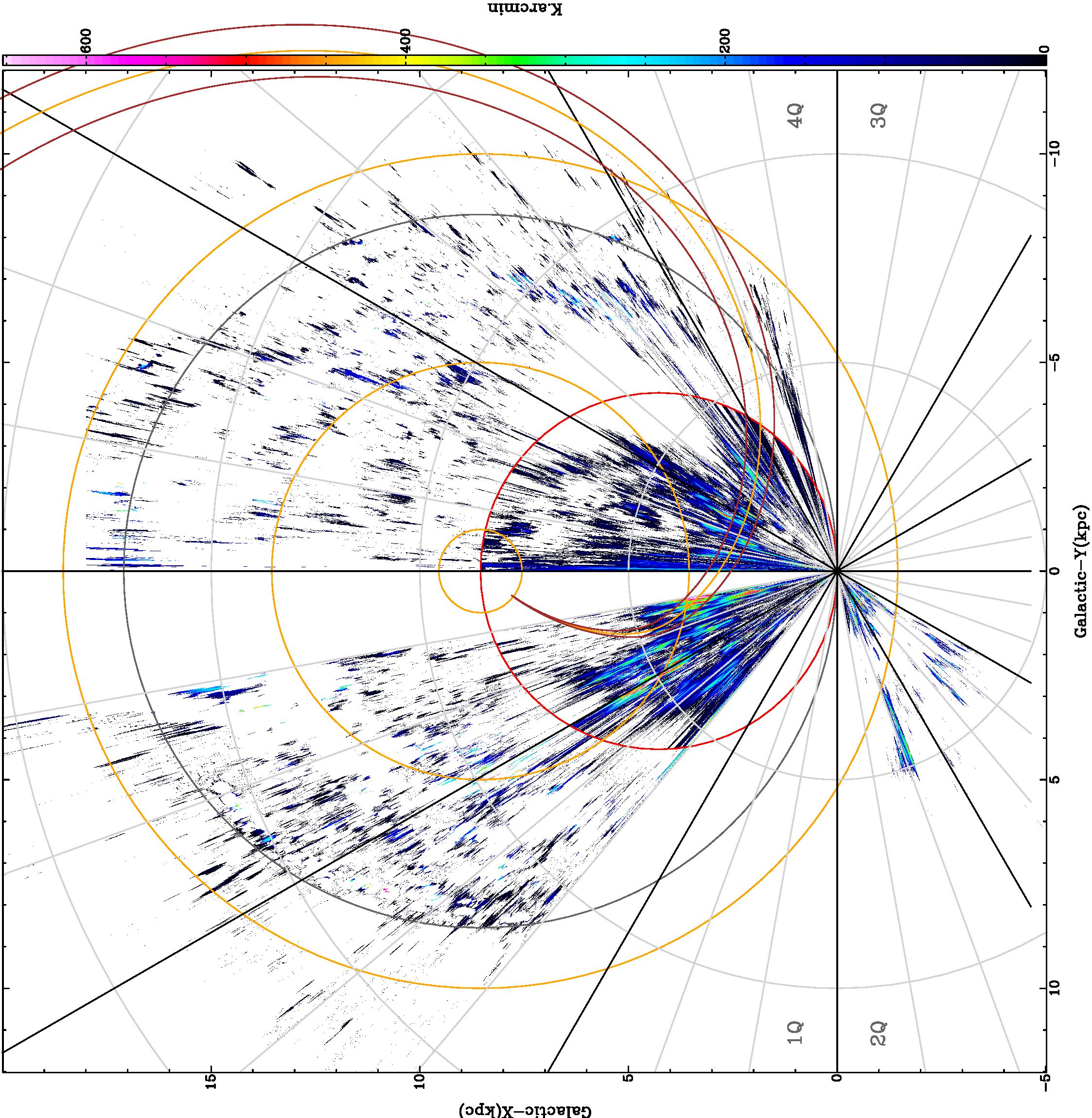} 
	\caption{\footnotesize Same WFTN zeroth latitude-moment \tco\ intensity data as in Figs.\,\ref{oldLVmodels} \& \ref{newhelio}, 
		converted to an \xy\ grid as in Fig.\,\ref{xysmall} (top panel) but zoomed out to show the whole transformation.  Here, 
		heliocentric distance contours are every 5\,kpc (grey); galactocentric distance contours are at 1, 5, 10\,kpc (orange) and 
		$R_0$ (the solar circle, black).}
	\label{wftn-xy0}
\end{figure*}

\vspace{-1mm}
Full Cartesian maps of the entire available WFTN coverage are shown here, similar to the zoom-ins of Figure \ref{xysmall}, \ref{smalldust}, and \ref{zdiff}, but as far as the data extend in each moment map.

The data behind these figures are also available at \texttt{https://gemelli.spacescience.org/$\sim$pbarnes/ research/codex/datafiles/} as downloadable compressed FITS files, listed below (file sizes are approximate).  These files constitute Data Release 1 (DR1) and the first 12 ``pages'' of the CODEX.

\textbullet\ Longitude-velocity diagrams: \hspace{84mm}{wftn.codex.lv0.fits.gz\,(51\,Mb)} \\ \rightline{ftn.codex.lv1.fits.gz (50 Mb)} \rightline{ftn.codex.lv2.fits.gz (49 Mb)}

\textbullet\ Longitude-distance\,diagrams: \\ \rightline{wftn.bvlg.codex.ld0.fits.gz\,(67\,Mb)} \\ \rightline{ftn.bvlg.codex.ld1h.fits.gz (61 Mb)} \rightline{ftn.bvlg.codex.ld2h.fits.gz (60 Mb)}

\textbullet\ Cartesian deprojections, where the axis order is y,x,z so that the Galactic Centre appears above the Sun on the page or screen, as displayed in the Figures: \hspace{60mm}{wftn.bvlg.codex.YX0.fits.gz (10 Mb)} \rightline{ftn.bvlg.codex.YX1h.fits.gz (9 Mb)} \rightline{ftn.bvlg.codex.YX2h.fits.gz (9 Mb)}

\textbullet\ Dust-related maps: \hspace{95mm}Dust.codex.YX0.fits.gz (60 Mb) \\ \rightline{Dust.codex.YX1h.fits.gz (8 Mb)} \rightline{GasMinusDust.codex.YX1h.fits.gz (8 Mb)}
\vspace{0mm}

\begin{figure*}
	\includegraphics[angle=-90,width=179mm]{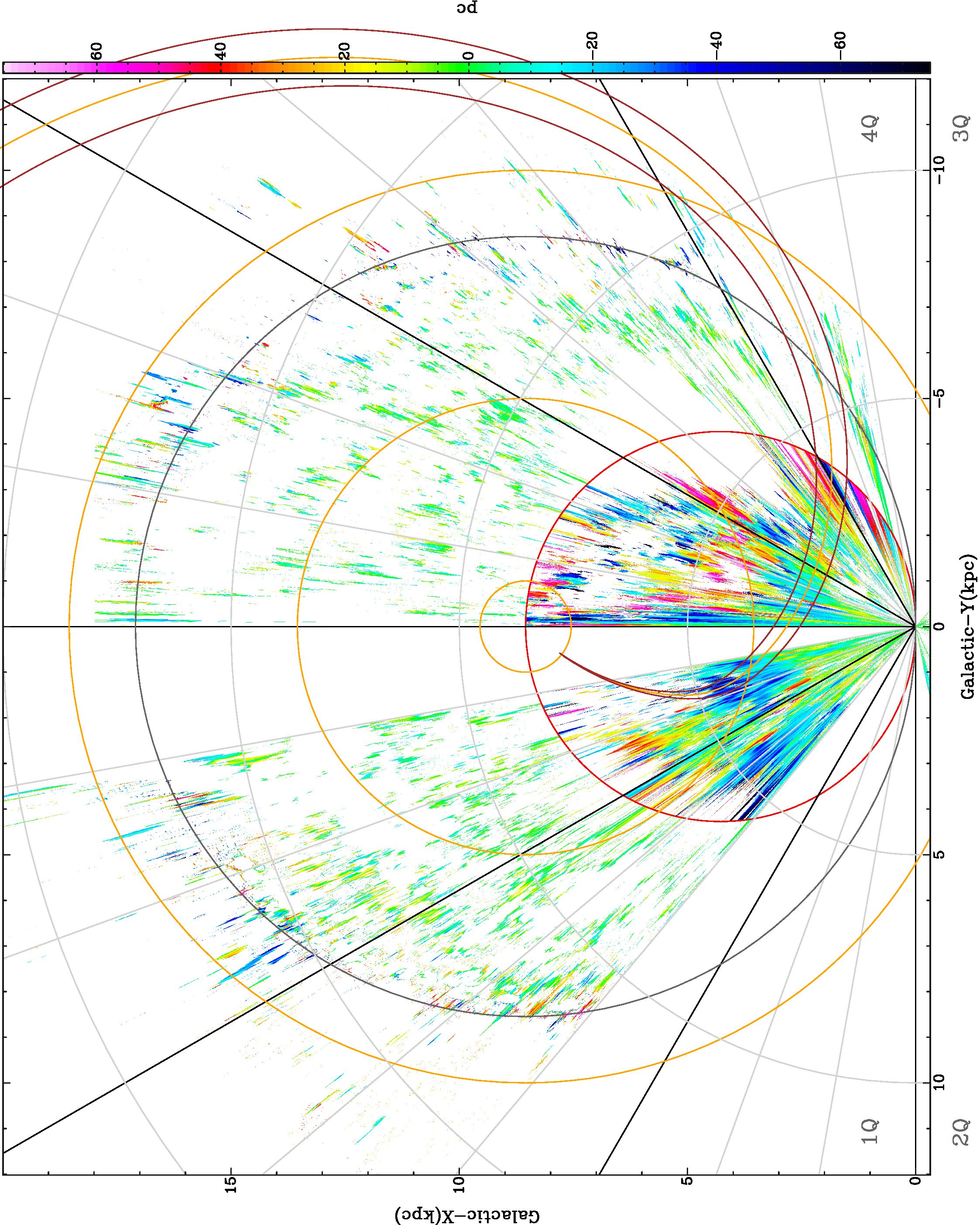} 
	\caption{\footnotesize As in Fig.\,\ref{wftn-xy0} but for the first latitude-moment $\bar{z}$ \lv\ maps from FTN data reprojected onto \xy, 
		and the angular latitude units converted to physical heights.}
	\label{ftn-xy1}
\end{figure*}

\begin{figure*}
	\includegraphics[angle=-90,width=179mm]{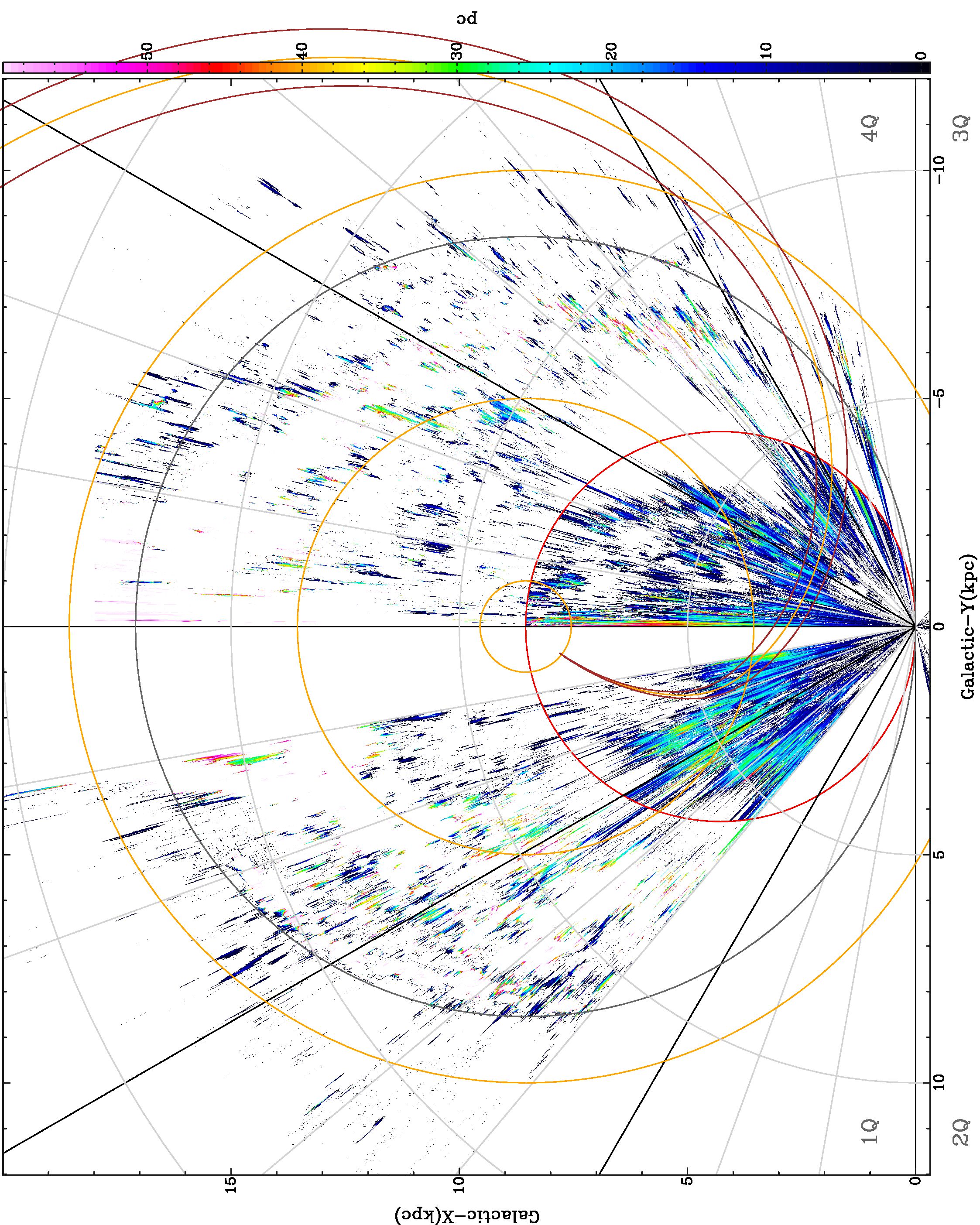} 
	\caption{\footnotesize As in Fig.\,\ref{wftn-xy0} but for the second latitude-moment $\sigma_{z}$ \lv\ maps from FTN data reprojected 
		onto \xy, and the angular latitude units converted to physical thicknesses.}
	\label{ftn-xy2}
\end{figure*}

\clearpage
\begin{figure*}
	\hspace{-1.5mm}\includegraphics[angle=0,width=179mm]{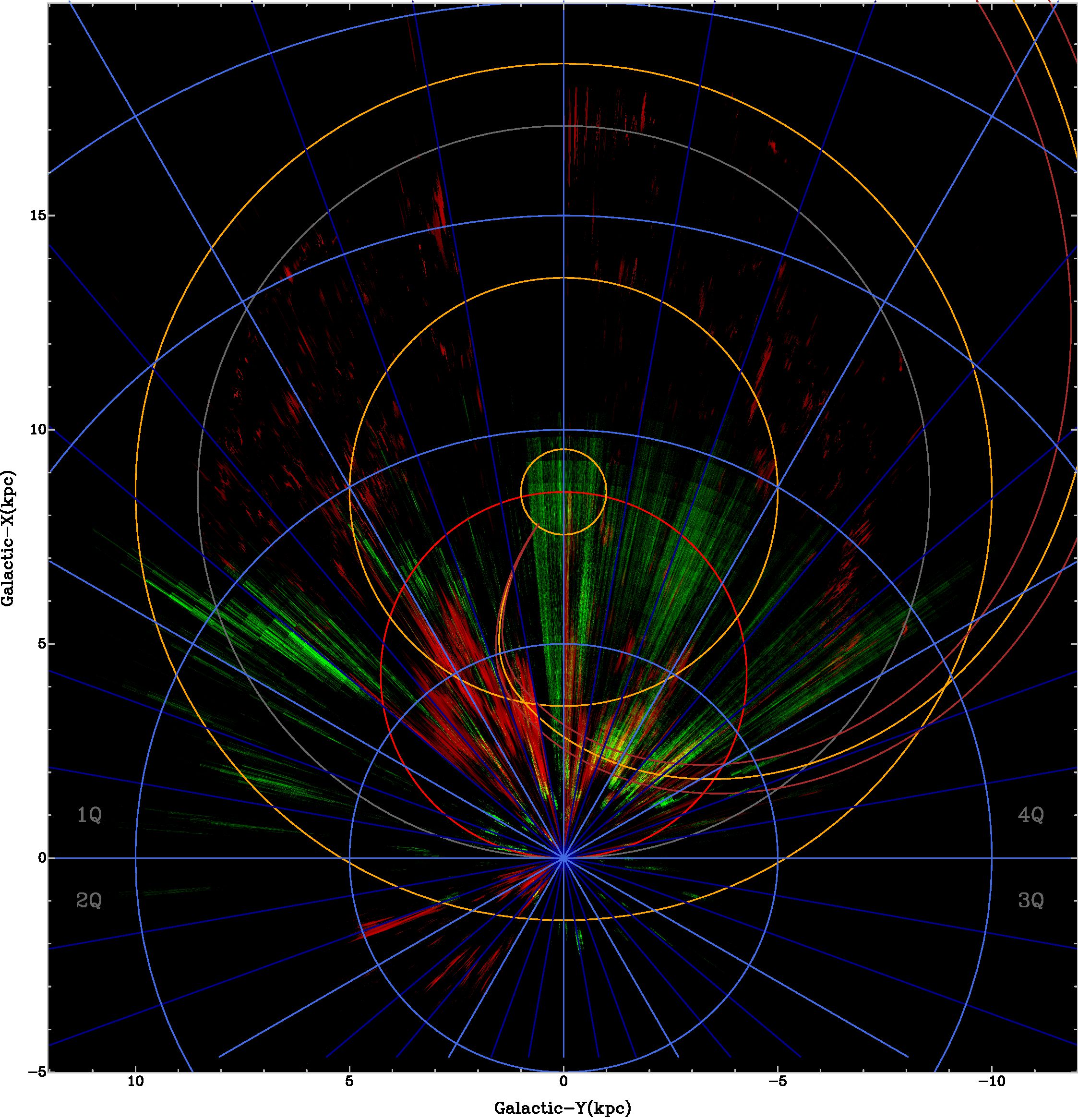} 
	\caption{\footnotesize Overlay of molecular gas and dust \xy\ maps as in Fig.\,\ref{smalldust}, but zoomed out to show the whole transformation.  
		Here, the blue heliocentric distance contours are every 5\,kpc; other features are the same as in Fig.\,\ref{smalldust}.
	\label{wftn-dust}}
\end{figure*}

\begin{figure*}
	\includegraphics[angle=-90,width=179mm]{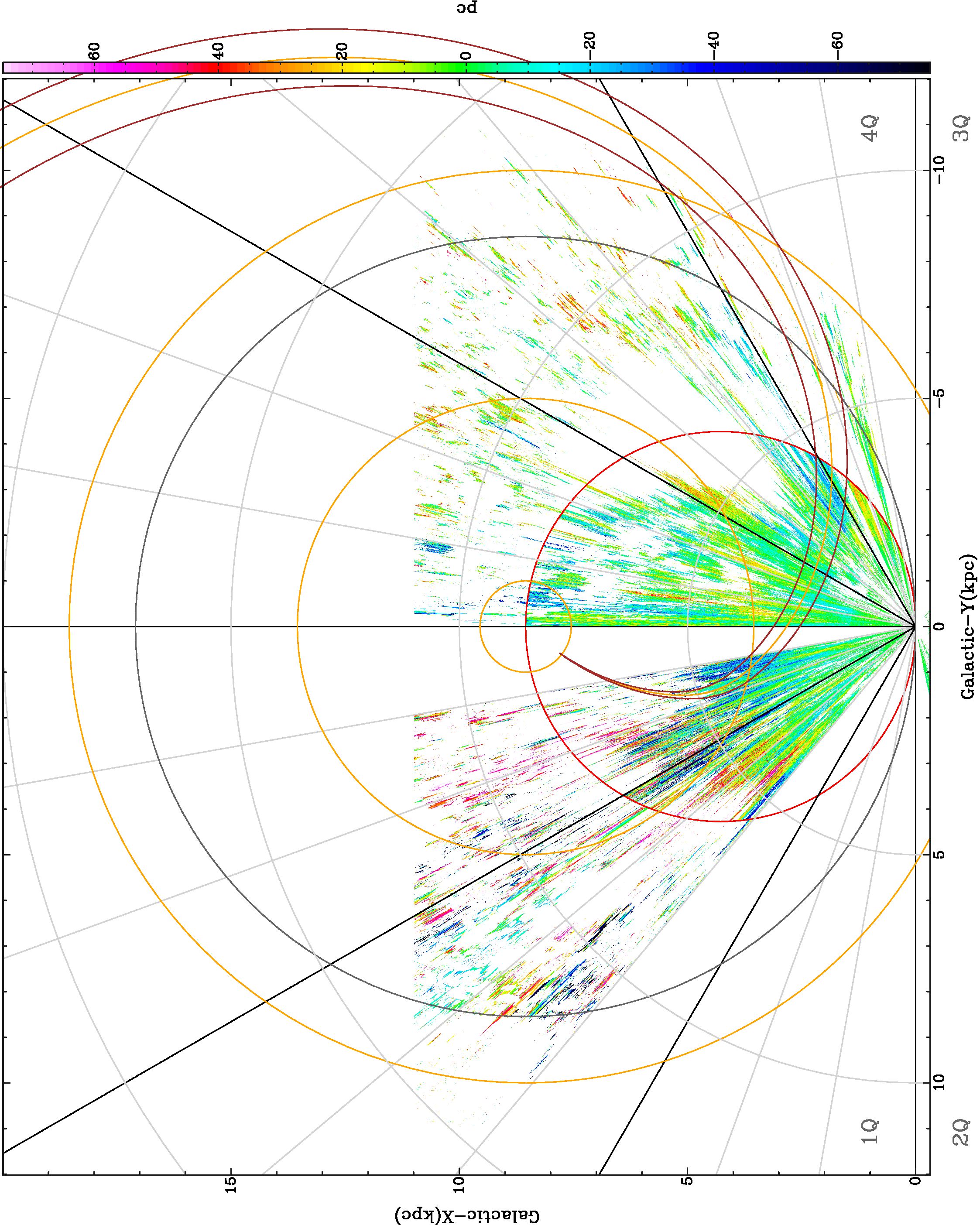}
	\caption{\footnotesize As in Fig.\,\ref{ftn-xy1} but for the first latitude-moment $\bar{z}$ of the \citet{z25} dust cube, limited to $z$ within the 
		corresponding angular latitude limits ($\pm1$\degree) of the FTN data, and to where the \xy\ pixels are not blanked in 
		Fig.\,\ref{ftn-xy1}.  For simplicity, the \citet{r19} spiral arms have also been omitted here.}
	\label{dust-xy1}
\end{figure*}

\section{CfA Survey Moment Maps}\label{cfaviews}

To illustrate the effects of their wider latitude coverage, deprojected moment maps of the CfA survey \citep{dht01} are shown here in the same manner as Appendix \ref{wideview}, using the same kinematic model and near/far disambiguation as described in this paper.  The main features of relevance to the various discussions herein include the following. 

\begin{figure*}
	\includegraphics[angle=0,width=179mm]{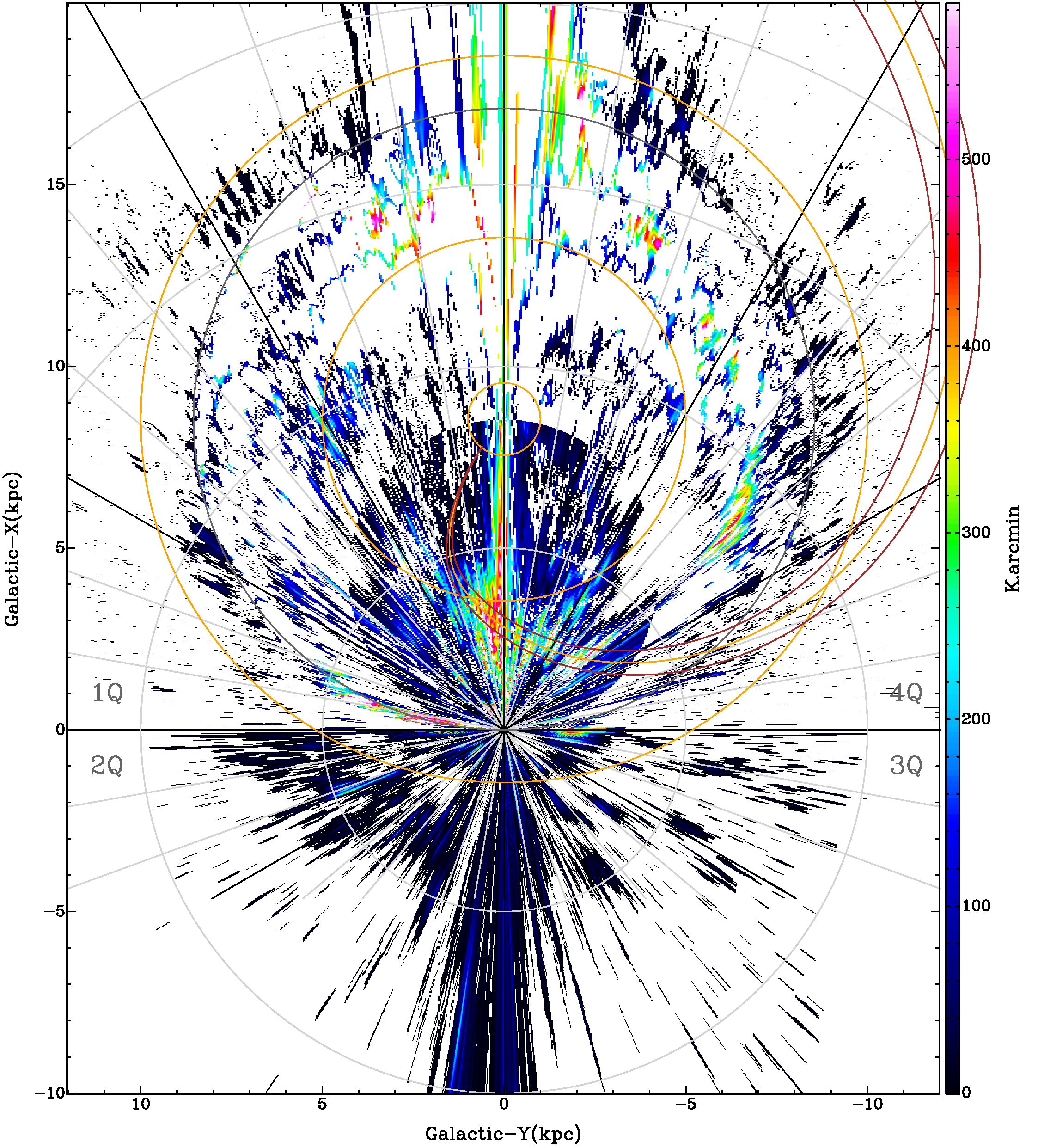}	
	\caption{\footnotesize Latitude-integrated \tco\ intensity as in Fig.\,\ref{wftn-xy0} but now from the CfA survey \citep{dht01}.}
	\label{cfa-xy0}
\end{figure*}

The RSA was fit exclusively to align with the deprojected features in the FUGIN+ThrUMMS \xy\ maps: the fact that it {\bf {\em also}} aligns extremely well with the maximum latitude-integrated \tco\ intensity from the CfA data is rather impressive.  The only additional high-\icob\ feature in Figure \ref{cfa-xy0} seems to lie close to the Solar Circle, i.e., near $l$ $\approx$ +80\degree\ and --90\degree, apparently due to that survey's much wider latitude coverage ($\pm$5\degree) than in the FUGIN+ThrUMMS data ($\pm$1\degree).  If this represents a coherent spiral feature missed by the higher-resolution surveys, its pitch angle ($\sim$5--10\degree) is much smaller than the RSA's, and would be suggestive of the ``missing'' Sgr-Car Arm except for the fact that this ``arm'' passes through the Sun's position, rather than lying $\sim$1 kpc inward of the Solar Circle.

\begin{figure*}
	\includegraphics[angle=0,width=179mm]{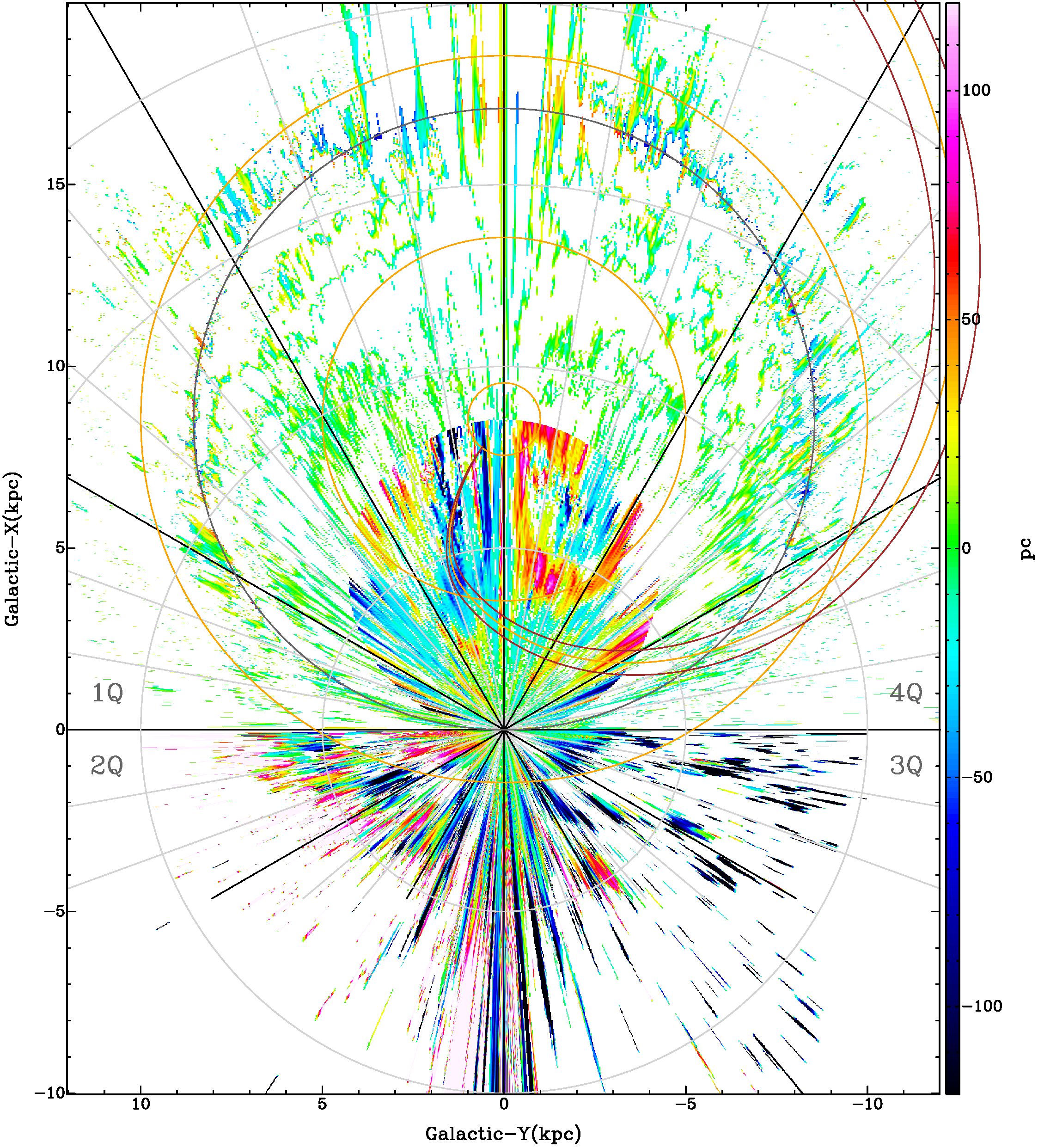}	
	\caption{\footnotesize Mean height of the dense gas layer as in Fig.\,\ref{ftn-xy1} but now from the CfA survey \citep{dht01}.}
	\label{cfa-xy1}
\end{figure*}

Figure \ref{cfa-xy1} underscores the very widespread nature of the ripples in both the inner and outer Galaxy.  
However, the inner ripple pattern is apparently coherent to somewhat larger scales than the outer ripples, 4--5\,kpc compared to 2--3\,kpc.  The well-known warp in the outer Galaxy is also evident in the 2+3Q CfA data, with deviations from the nominal Plane up to 300\,pc in either direction.

\begin{figure*}
	\includegraphics[angle=0,width=179mm]{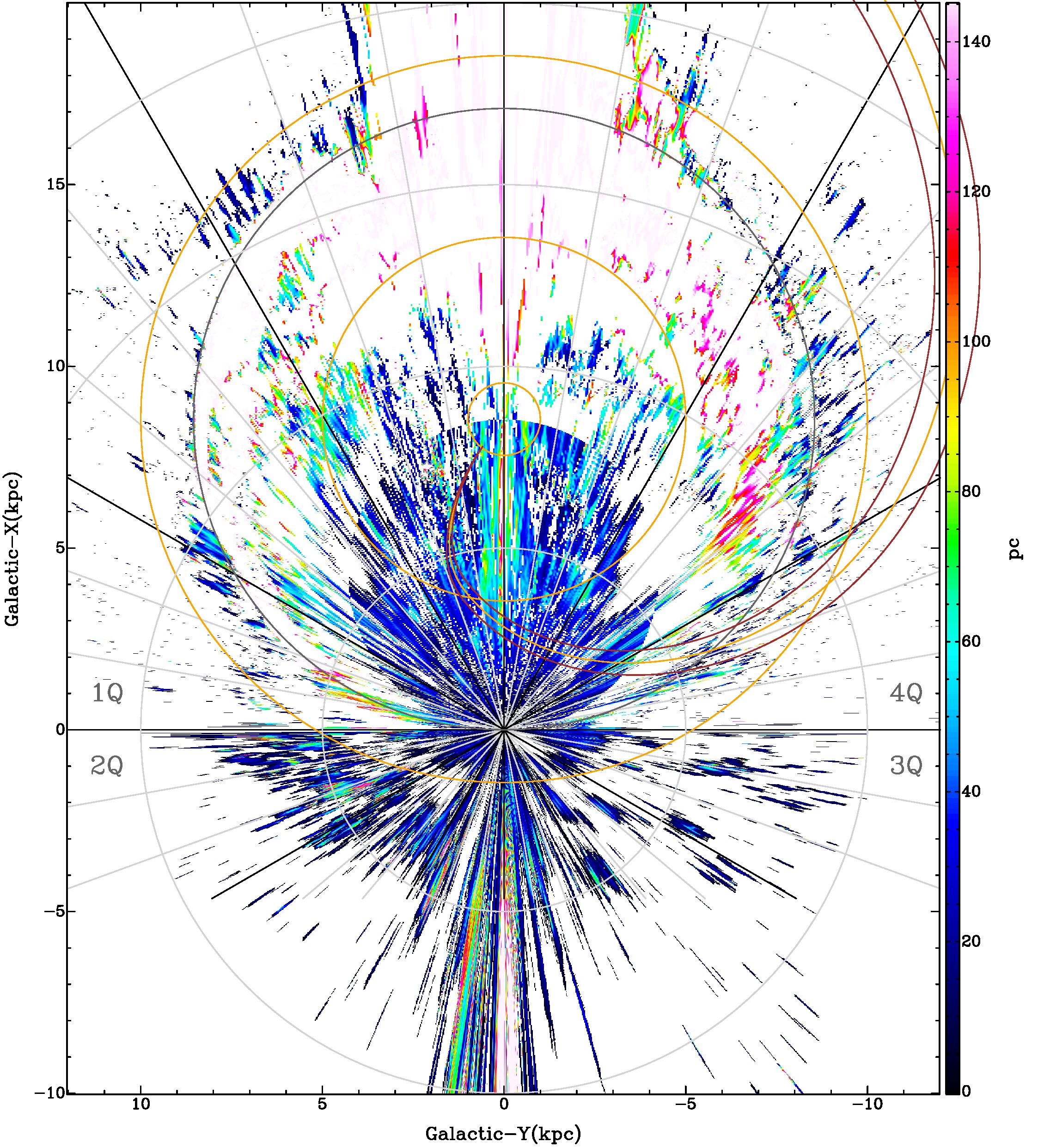}	
	\caption{\footnotesize Thickness of the molecular gas layer as in Fig.\,\ref{ftn-xy2} but now from the CfA survey \citep{dht01}.}
	\label{cfa-xy2}
\end{figure*}

Finally, the deprojected thickness $\sigma_{z}$ map (Fig.\,\ref{cfa-xy2}) reveals some variations across the disk, but this is more muted than in the higher-resolution data (Fig.\,\ref{ftn-xy2}).  In the inner Galaxy we find a modal thickness around 38\,pc, double the value found by \cite{b25}, but $\sigma_{z}$ becomes more variable again in the outer MW.

\bibliographystyle{aasjournal}




\end{document}